\documentclass[aip,amsmath,amssymb,reprint]{revtex4-1}

\usepackage{graphicx}
\usepackage{dcolumn}
\usepackage{bm}
\usepackage[T1]{fontenc}
\usepackage{xcolor}
\usepackage{float}
\usepackage{pgfplotstable}
\usepackage{array}
\usepackage{booktabs}
\usepackage[export]{adjustbox}
\usepackage{hyperref}
\usepackage{soul}
\usepackage[version=4]{mhchem}
\usepackage{amssymb}
\usepackage{braket}
\usepackage{xspace}

\newcommand{\ie}[0]{\textit{i.e.}, }
\newcommand{\eg}[0]{\textit{e.g.}, }
\newcommand{\cf}[0]{\textit{cf}.\ }
\newcommand{\via}[0]{\textit{via} }
\newcommand{\Li}{\ce{Li+}}
\newcommand{\Na}{\ce{Na+}}
\newcommand{\F}{\ce{F-}}
\newcommand{\Cl}{\ce{Cl-}}

\newcommand{\pII}[0]{\textsc{paper-ii}\xspace}
\usepackage{scalefnt}
\newcommand\smaller[2][0.85]{{\scalefont{#1}#2}}
\newcommand{\NENCI}{\textsc{nenci-}\smaller{2021}\xspace}
\newcommand{\DefineAuthor}[2]{%
  \expandafter\newcommand\csname #1note\endcsname[1]{%
    \textbf{\textcolor{#2}{\textbf{#1:} ##1}}}%
  \expandafter\newcommand\csname #1\endcsname[1]{
    \textbf{\textcolor{#2}{##1}}}
  \expandafter\newcommand\csname #1cancel\endcsname[1]{%
    \textbf{\textcolor{#2}{\sout{##1}}}}%
  \expandafter\newcommand\csname #1change\endcsname[2]{%
    \textbf{\textcolor{#2}{\sout{##1} ##2}}}%
  \newenvironment{#1text}{\color{#2}}{\color{black}}
}
\definecolor{dartmouthgreen}{rgb}{0.05, 0.5, 0.06}
\DefineAuthor{RAD}{red}
\DefineAuthor{BGE}{blue}
\DefineAuthor{ZMS}{magenta}
\DefineAuthor{PTJ}{orange}
\DefineAuthor{GEN}{dartmouthgreen}

\begin{document}

\title{NENCI-2021 Part I: A Large Benchmark Database of Non-Equilibrium Non-Covalent Interactions Emphasizing Close Intermolecular Contacts}

\author{Zachary M. Sparrow}
\thanks{These authors contributed equally to this work.}
\author{Brian G. Ernst}
\thanks{These authors contributed equally to this work.}
\author{Paul T. Joo}
\author{Ka Un Lao}
\author{Robert A. DiStasio Jr.}
\email{distasio@cornell.edu}
\affiliation{Department of Chemistry and Chemical Biology, Cornell University, Ithaca, NY 14853 USA}
\date{\today}

\begin{abstract}
\noindent In this work, we present \NENCI, a benchmark database of approximately ${8,000}$ non-equilibrium non-covalent interaction energies for a large and diverse selection of intermolecular complexes of biological and chemical relevance.
To meet the growing demand for large and high-quality quantum mechanical data in the chemical sciences, \NENCI starts with the $101$ molecular dimers in the widely used S66 and S101 databases, and extends the scope of these works by: (\textit{i}) including $40$ cation- and anion-$\pi$ complexes, a fundamentally important class of non-covalent interactions (NCIs) that are found throughout nature and pose a substantial challenge to theory, and (\textit{ii}) systematically sampling all $141$ intermolecular potential energy surfaces (PES) by \textit{simultaneously} varying the intermolecular distance and intermolecular angle in each dimer.
Designed with an emphasis on close contacts, the complexes in \NENCI were generated by sampling seven intermolecular distances along each PES (ranging from $0.7\times\mathrm{-}1.1\times$ the equilibrium separation) as well as nine intermolecular angles per distance (five for each ion-$\pi$ complex), yielding an extensive database of ${7,763}$ benchmark intermolecular interaction energies ($E_{\rm int}$) obtained at the CCSD(T)/CBS level of theory. 
The $E_{\rm int}$ values in \NENCI span a total of $225.3$~kcal/mol, ranging from $-38.5$~kcal/mol to $+186.8$~kcal/mol, with a mean (median) $E_{\rm int}$ value of $-1.06$~kcal/mol ($-2.39$~kcal/mol).
In addition, a wide range of intermolecular atom-pair distances are also present in \NENCI, where close intermolecular contacts involving atoms that are located within the so-called van der Waals envelope are prevalent---these interactions in particular pose an enormous challenge for molecular modeling and are observed in many important chemical and biological systems.
A detailed SAPT-based energy decomposition analysis also confirms the diverse and comprehensive nature of the intermolecular binding motifs present in \NENCI, which now includes a significant number of primarily induction-bound dimers (\eg cation-$\pi$ complexes).
\NENCI thus spans all regions of the SAPT ternary diagram, thereby warranting a new four-category classification scheme that includes complexes primarily bound by electrostatics (${3,499}$), induction ($700$), dispersion (${1,372}$), or mixtures thereof (${2,192}$).
A critical error analysis performed on a representative set of intermolecular complexes in \NENCI demonstrates that the $E_{\rm int}$ values provided herein have an average error of $\pm \, 0.1$~kcal/mol, even for complexes with strongly repulsive $E_{\rm int}$ values, and maximum errors of $\pm \, 0.2\mathrm{-}0.3$~kcal/mol (\ie approximately $\pm \, 1.0$~kJ/mol) for the most challenging cases.
For these reasons, we expect that \NENCI will play an important role in the testing, training, and development of next-generation classical and polarizable force fields, density functional theory (DFT) approximations, wavefunction theory (WFT) methods, as well as machine learning (ML) based intra- and inter-molecular potentials.
\end{abstract}

\maketitle

\section{\label{sec:intro} Introduction}

With tunable strengths situated between thermal fluctuations and covalent bonds, non-covalent interactions (NCIs) are ubiquitous in nature and play a critical role in determining the structure, stability, and function in a number of systems throughout chemistry, biology, physics, and materials science.~\cite{Langbein_vdw_theory_1974,vdw_Handbook_2005,Kaplan_Intermol_Picture_Comp_2006,Intermolecular_Surface_Forces_3rd_2011,Stone_Intermolecular_Forces}
One particularly illustrative example is the famous DNA double helix, whose structure is stabilized by a complex network of hydrogen bonds and $\pi$-$\pi$ stacking interactions between constituent nucleobases.
In organic synthesis and biochemistry, many catalysts and enzymes function by leveraging NCIs to position/orient substrates for the ensuing reaction and/or stabilize critical points along the reaction pathway, \eg ion-$\pi$ interactions can stabilize intermediates and transition states with excess charge.~\cite{Squalene_Cyclase_1997,anion-pi_catalysis_zhao_2013,anion-pi_catalysis_2014,cation-pi_catalysis_2016,account_of_anion-pi_catalysis_2018,yamada_cation-pi_org_synth_review_2018}
Over the past two decades, NCIs have garnered critical recognition throughout the chemical sciences, and have now become an integral part of ``chemical intuition'' when rationalizing complex chemical structures and/or processes as well as designing  molecular systems (\eg catalysts) for optimal performance and/or novel applications.
In this regard, there are quite a number of NCI-based applications actively under investigation, ranging from crystal engineering (where hydrogen and halogen bonding are used to direct molecular assembly)~\cite{engineering_by_halogen_bonds_2007,halogen_bond_crystal_eng_anion_template_2009} and artificial molecular machines (where $\pi$-$\pi$ stacking, hydrogen bonding, and dispersion/van der Waals (vdW) forces are leveraged to control complex motion at the nanoscale)~\cite{molecular_tweezers_1978,stoddart_molecular_shuttles_1994,molecular_machines_1998,artificial_molecular_machines_2000} to drug discovery (where candidate molecules are selected and screened based on specific NCIs present in the corresponding active site).~\cite{structure_based_discovery_docking_2009,hobza_reliable_drug_docking_2010,med_chem_guide_to_interactions_2010,Hobza_malonate_docking_w_experiment_2015}

Given their importance and prevalence, it is imperative that there exists a suite of computational methods that can provide an accurate, reliable, and computationally efficient description of NCIs for systems ranging from small gas-phase molecular dimers to the complex tertiary and quaternary structure of proteins in solvent.~\cite{protein_folding_50_years}
To meet these goals, a number of computational techniques have been developed over the past century, including (but not limited to): model intermolecular potentials (\eg Lennard-Jones),~\cite{LJ_potential_from_eq_of_state_1924} classical and polarizable force fields,~\cite{AMOEBA_status_2010,polarizable_forcefield_review_2016} density functional theory (DFT) approximations with corrections for dispersion/vdW interactions,~\cite{Grimme_disp_corrected_DFT_review_2011,vdw_functional_review_2015,DiStasio_vdw_review_2017} efficient (\ie linear scaling) algorithms for highly accurate wavefunction theory (WFT) methods,~\cite{dlpno-ccsd_2013,dlpno-ccsdpt_2013} and more recently, the large and rapidly growing suite of machine learning (ML) based approaches.~\cite{Rupp_ML_atomization_2012,bypass_KS_ML_2017,Tkatchenko_SchNet_2018,Bereau_ML_PT_NCI_2018,Sherrill_ML_NCI_SAPT_2020}
During this time, such computational methods have enjoyed tremendous success and made critical contributions to a number of different fields, \eg identifying promising pharmaceutical molecules,~\cite{structure_based_discovery_docking_2009,hobza_reliable_drug_docking_2010,med_chem_guide_to_interactions_2010,Hobza_malonate_docking_w_experiment_2015} predicting (meta-)stable molecular crystal polymorphs,~\cite{polymorph_blind_test_report_2016,Hoja_polymorphs_2019} and elucidating supercritical behavior in high-pressure liquid hydrogen,~\cite{ceriotti_liquid_hydrogen_2020} to name a few.
However, we would argue that the next generation of theoretical approaches for describing NCIs would tremendously benefit by addressing the following challenges in an accurate, reliable, and computationally efficient manner: 
(\textit{i}) the need to describe NCIs in large molecular and condensed-phase systems (\ie collective many-body effects, solvation/solvent effects, simultaneous treatment of short-, intermediate-, and long-range NCIs on the same footing);~\cite{grimme_s12l_2012,tkatchenko_mbd_2012,distasio2012collective,sedlak_l7_2013,goldey2014shared,ambrosetti2014hard} 
(\textit{ii}) the need to describe the diverse types of NCIs on the same footing (\ie similar performance for hydrogen bonding, $\pi$-$\pi$ stacking, dispersion, ion-$\pi$ interactions, etc);~\cite{DiStasio_SCS-MI-MP2_2007,Hobza_MP2_basis_2007,Hobza_MP2_Basis_Performance_2012,DNA_IonPi_2020} and
(\textit{iii}) the need to describe NCIs in equilibrium and non-equilibrium systems on the same footing (\ie similar performance across entire potential energy surfaces (PES)).~\cite{wang_s101x7_2015,gould2018dispersion}

An essential part of developing next-generation theoretical methods for describing NCIs involves testing and/or training new approximations against highly accurate benchmark data.
For the increasingly popular suite of ML-based models---which require large amounts of high-quality data to learn the quantum mechanics underlying NCIs---such reference data is of critical importance.
However, such benchmark non-covalent/non-bonded interaction (or binding) energies ($E_{\rm int}$) are seldom experimentally available, especially for large/complex systems and non-equilibrium configurations. 
Instead, one usually relies on quantum chemical and/or quantum Monte Carlo methods (\ie WFT methods) to obtain highly accurate and systematically improvable $E_{\rm int}$ values for benchmarking and training purposes.~\cite{benchmark_theory_on_theory_2017}
On the WFT side, coupled cluster theory including single, double, and perturbative triple excitations in conjunction with an extrapolation to the complete basis set limit (CCSD(T)/CBS) has long been considered the \textit{de facto} ``gold standard'' for generating accurate $E_{\rm int}$ data for small- and medium-sized organic molecules, and has therefore been used to generate a number of seminal benchmark databases for NCIs.~\cite{Hobza_ccsdt_gold_standard_2013,Sherrill_Silver_Bronze_NCI_2014}
One of the first of these databases, the so-called S22 database,~\cite{jurecka_s22_2006,Sherrill_DeltaCC_extrapolation_2011} includes $22$ CCSD(T)-quality $E_{\rm int}$ values for a set of small-/medium-sized biologically-relevant intermolecular complexes (comprised of \{\ce{C}, \ce{H}, \ce{O}, \ce{N}\}) in their respective optimized (equilibrium) geometries, and was designed to cover a number of different intermolecular binding motifs (\ie single and double hydrogen bonds, dipole-dipole interactions, $\pi$-$\pi$ stacking, dispersion, \ce{C}--\ce{H}$\cdots\pi$, etc).
Following the success and widespread use of S22 in the testing and parameterization of many theoretical methods for describing NCIs, the amount of benchmark-quality $E_{\rm int}$ data was substantially increased with the introduction of the S66 database (which includes $66$ equilibrium intermolecular complexes of similar size and composition to that found in S22) as well as extensions thereof to include complexes with non-equilibrium intermolecular distances (along a series of dissociation curves in S22x5~\cite{Hobza_S22x5_Method_Comparison_2010} and S66x8~\cite{rezac_s66_2011,JMartin_S66x8_2016}) and non-equilibrium intermolecular angles (at the equilibrium distance in S66a8).~\cite{rezac_s66a8_2011}
During the same time, other benchmark NCI databases were constructed to reflect the diverse number of NCI types (or binding motifs) found in: halogen-containing systems (X40x10),~\cite{rezac_x40_2012} nucleobase dimers (ACHC),~\cite{parker_achc_2015} charge transfer complexes (CT),~\cite{truhlar_ct_2005} alkane dimers (ADIM6),~\cite{tsuzuki_adim6_2006} large molecular dimers (L7),~\cite{sedlak_l7_2013} host-guest complexes (S12L),~\cite{grimme_s12l_2012} halogen-bonded systems (XB18),~\cite{kozuch_xb18_2013} sulfur-containing systems (SULFURx8),~\cite{Parks_sulfurx8_2012} and many more.~\cite{Sherrill_DeltaCC_extrapolation_2011,hobza_ionichb_2012,sherrill_hsg_2011,Hobza_ccsdt_gold_standard_2013,Hobza_Extended_A24_2015,sherrill_ssi_2016,Grimme_DFT_Zoo_2017}
Along similar lines, there are also NCI databases based on a symmetry-adapted perturbation theory (SAPT) decomposition of $E_{\rm int}$ into components (\ie electrostatics, exchange, induction, and dispersion), which have been used to train force fields for molecular dynamics simulations.~\cite{schmidt_sapt-ff_2013,schmidt_universal_approach_to_ff_2014,vandenbrande_medff_2017}
Of particular interest here is the S101x7 database,~\cite{wang_s101x7_2015} which starts with the molecular dimers in S66 and expands this set to include $35$ additional biologically-relevant complexes containing halogens (\ie \ce{F}, \ce{Cl}, \ce{Br}) and second-row elements (\ie \ce{S} and \ce{P}), as well as additional intermolecular complexes involving charged systems and/or water.
Like the S66x8 database, S101x7 also includes complexes with non-equilibrium intermolecular distances by computing SAPT-based $E_{\rm int}$ values for select points along each intermolecular PES; in the S101x7 case, these seven points ranged from $0.7\times\mathrm{-}1.1\times$ the equilibrium intermolecular separation in an effort to better capture short-range charge penetration effects.~\cite{wang_s101x7_2015}

While such existing databases are growing in size, most are still relatively small (containing $\lesssim 500$ interaction energies), making them insufficient for the rapidly growing field of ML-based intra-/inter-molecular potentials.
While composite databases~\cite{Grimme_DFT_Zoo_2017,ACCDB_2019} can be considerably larger, the accuracy and reliability of such compiled data is inconsistent and can potentially be a source of both random and systematic error. 
Due to the high computational cost of generating benchmark $E_{\rm int}$ values for large systems, most existing databases (with the exception of L7~\cite{sedlak_l7_2013} and S12L~\cite{grimme_s12l_2012}) have been limited to small-to-medium organic/biological molecules (usually containing $< 20$ atoms); as a consequence, many of these databases do not capture the collective nature of NCIs (\ie many-body effects, solvation/solvent effects, NCIs across multiple length scales) present in large/complex molecules and condensed-phase systems. 
In addition, most existing databases have focused on common intermolecular binding motifs such as hydrogen and halogen bonding, $\pi$-$\pi$ stacking, dipole-dipole interactions, dispersion, and \ce{C}--\ce{H}$\cdots\pi$ interactions, while other important binding motifs (like cation- and anion-$\pi$ interactions) have been largely underrepresented.
As such, these databases tend to include intermolecular complexes that are primarily bound by electrostatics, dispersion, or a mixture thereof, but have not included intermolecular complexes that are primarily induction-bound.
Furthermore, prior databases (\eg S22x5, S66x8, S66a8, S101x7) primarily focused on the equilibrium geometry and a \textit{single} displacement from the equilibrium geometry (\ie scaling the intermolecular distance or rotating one monomer), but very few have explored wider swaths of the intermolecular PES.
In this regard, most databases have also only slightly touched upon close intermolecular contacts (\ie the short-range and often repulsive sector of the intermolecular PES), although there are some examples where such short-range considerations have been incorporated (\eg S101x7~\cite{wang_s101x7_2015} and R160x6~\cite{R160x6_repulsive_contacts_2018}).
As a result, the performance of many theoretical methods for accurately and reliably describing NCIs in large and complex systems, for a diverse array of binding motifs, and across significant portions of the intermolecular PES is simply not well known.

Accurate and reliable descriptions of non-equilibrium NCIs at reduced intermolecular separations---where several strong and competing short-range intermolecular forces are at play---are important for a number of reasons and pose a substantial challenge to theory.
For instance, there are numerous examples throughout chemistry and chemical biology where close intermolecular contacts are either present at equilibrium or force the system to adopt a different configuration.~\cite{Exponential_Docking_2016,SLin_dual_electrocatalysis_2020} 
A striking example of this was recently observed when studying the enantioselectivity of sBOX catalysts, where a combination of attractive and repulsive NCIs are responsible for the enantio-determining \ce{C-CN} bond formation in chiral nitriles.~\cite{SLin_dual_electrocatalysis_2020}
Intermolecular close contacts also play a crucial role in the study of systems operating under high-pressure conditions, ranging from the microscopic structure of supercritical water~\cite{supercritical_water_2013} to the high-pressure synthesis of compounds with atypical compositions and novel properties~\cite{miao_high_pressure_chemistry_2020} as well as the search for high-$T_c$ superconducting materials.~\cite{Hoffmann_Ashcroft_high_pressure_high_Tc_superconductor_2017}
Theoretically speaking, SAPT decomposition studies~\cite{Sherrill_BZ_CP_2011} have shown that the intermolecular distance can have a profound influence on the absolute and relative magnitudes of the underlying $E_{\rm int}$ components (\ie electrostatics, exchange, induction, and dispersion), implying that the forces present in short-range non-equilibrium intermolecular complexes of small/simpler molecules can mimic those found in larger/more complicated systems at equilibrium separations.
Interestingly, this also suggests that training and/or testing theoretical methods on non-equilibrium configurations (particularly in the short-range) of small-to-medium molecular dimers can be used as a surrogate for describing the NCIs in a more diverse range of large (and possibly intractable) systems.
Since finite-temperature molecular dynamics and Monte Carlo simulations require a consistent treatment of the structures and energetics across the entire PES, an accurate and reliable treatment of non-equilibrium NCIs (including short-range as well as intermediate- and long-rang interactions) is of enormous importance for these applications as well.
However, the difficulties in obtaining such an accurate and reliable theoretical description of non-equilibrium NCIs across multiple length scales should also be emphasized.
For instance, the long-range sector of the intermolecular PES requires a balanced description of both electrostatics and dispersion, and this can be particularly challenging when dealing with NCIs that also include charged species and/or molecules with substantial multipole moments.
For larger intermolecular separations, intermolecular energies (and forces) tend to be small, which provides additional challenges when trying to describe points along the PES on the same footing.
At reduced intermolecular separations, the increased amount of orbital (or density) overlap between monomers gives rise to a complex interplay between strongly attractive and strongly repulsive intermolecular forces (\eg charge transfer and penetration, Pauli repulsion, many-body exchange-correlation effects, etc), and an error when describing any one of these components can lead to disastrous results.~\cite{SAPT_Review_Jeziorski_1994,KaUnLao_1exch_breakdown_2012,wang_s101x7_2015}
For such short-range non-equilibrium NCIs, the performance of the current suite of theoretical methods is still an open question, and a number of studies have reported higher errors for repulsive intermolecular contacts.~\cite{Sherrill_DeltaCC_extrapolation_2011,KaUnLao_1exch_breakdown_2012,mardirossian_30_years_dft_2017,R160x6_repulsive_contacts_2018}
In this regime, even the suitability of high-level WFT-based approaches for generating benchmark $E_{\rm int}$ data is still largely unresolved as such approaches suffer from issues related to the use of incomplete basis sets (\ie basis set incompleteness and superposition errors) in conjunction with an approximate treatment of electron correlation effects (including questions regarding the reliability of perturbative expansions).

In this work, we directly address the aforementioned challenges needed for training, testing, and developing next-generation theoretical approaches for describing NCIs by introducing \NENCI, a benchmark database of approximately ${8,000}$ \textbf{N}on-\textbf{E}quilibrium \textbf{N}on-\textbf{C}ovalent \textbf{I}nteraction energies for a diverse selection of $141$ molecular dimers of biological and chemical relevance.
Starting with the $101$ dimers in the S101~\cite{wang_s101x7_2015} (and hence S66~\cite{rezac_s66_2011}) databases, which contain a diverse set of intermolecular binding motifs (\ie single and double hydrogen bonds, halogen bonds, ion-dipole and dipole-dipole interactions, $\pi$-$\pi$ stacking, dispersion, \ce{X}--\ce{H}$\cdots\pi$) as well as a large number of molecular dimers involving water (which represents a crucial first step towards generating benchmark $E_{\rm int}$ values in aqueous environments), \NENCI extends the scope of these seminal works in two directions.
For one, \NENCI includes $40$ cation- and anion-$\pi$ complexes, a fundamentally important and particularly strong class of NCIs that are primarily induction-bound~\cite{DNA_IonPi_2020} and characterized by equilibrium $E_{\rm int}$ values which are typically larger in magnitude than hydrogen bonds and salt bridges.
As such, an accurate and reliable description of ion-$\pi$ interactions poses substantial difficulties for theory, and their inclusion in \NENCI directly addresses the challenge of simultaneously describing diverse NCI types on the same footing (\ie point (\textit{ii}) above).
Secondly, \NENCI also includes an extensive and systematic sampling of equilibrium and non-equilibrium configurations on each of the $141$ intermolecular PES by \textit{simultaneously} varying the intermolecular distance and intermolecular angle in each dimer.
Designed with an emphasis on close intermolecular contacts, the complexes in \NENCI were generated by sampling seven intermolecular distances (ranging from $0.7\times\mathrm{-}1.1\times$ the equilibrium separation) as well as nine intermolecular angles per distance (five for each ion-$\pi$ complex), yielding an extensive database of ${7,763}$ benchmark $E_{\rm int}$ values obtained at the CCSD(T)/CBS level of theory. 
In doing so, \NENCI directly addresses the challenges of describing the collective nature of NCIs in large/complex systems (\ie point (\textit{i})) and simultaneously describing NCIs in equilibrium and non-equilibrium systems on the same footing (\ie point (\textit{iii})). 
The $E_{\rm int}$ values in \NENCI span a total of $225.3$~kcal/mol, ranging from $-38.5$~kcal/mol (corresponding to the strongly attractive \Li$\cdots$Benzene ion-$\pi$ complex) to $+186.8$~kcal/mol (corresponding to a strongly repulsive $\text{DMSO}\cdots\text{DMSO}$ complex that has been scaled to $0.7\times$ the equilibrium intermolecular separation and rotated to a non-equilibrium angle), with a mean (median) $E_{\rm int}$ value of $-1.06$~kcal/mol ($-2.39$~kcal/mol).
A detailed SAPT-based energy decomposition analysis demonstrates the diverse and comprehensive nature of \NENCI, which spans all regions of the corresponding ternary diagram and includes intermolecular binding motifs primarily bound by electrostatics (${3,499}$), induction ($700$), dispersion (${1,372}$), or mixtures thereof (${2,192}$). 
A critical error analysis performed on a representative set of intermolecular complexes in \NENCI demonstrates that the $E_{\rm int}$ values provided herein at the CCSD(T)/CBS level have an average error of $\pm \, 0.1$~kcal/mol, even for complexes with strongly repulsive $E_{\rm int}$ values, and maximum errors of $\pm \, 0.2\mathrm{-}0.3$~kcal/mol (\ie approximately $\pm \, 1.0$~kJ/mol) for the most challenging cases.
Designed to meet the growing demand for large and high-quality quantum mechanical data in the chemical sciences, we expect that \NENCI will be an important resource for testing, training, and developing next-generation force fields, DFT approximations, WFT methods, and ML-based intra-/inter-molecular potentials.

The remainder of this manuscript is organized as follows.
Section II describes the construction of \NENCI, including the selection of molecular dimers, generation of equilibrium and non-equilibrium intermolecular complexes, a detailed description of the employed computational protocol, and a guide to obtaining the database.
Section III discusses the properties of \NENCI, including a statistical analysis of the intermolecular interaction energies and closest intermolecular contacts, an SAPT-based energy decomposition analysis of the intermolecular binding motifs, as well as a critical assessment of the error in the benchmark $E_{\rm int}$ values provided herein.
The manuscript ends with some brief conclusions and future directions in Section IV.
In a follow-up to this work,~\cite{NENCI_part_II} many popular WFT and DFT methods are explicitly tested on the \NENCI database, where it is shown that there is a nearly universal increase in error when describing the repulsive wall of the  intermolecular PES and that ion-$\pi$ complexes can be quite challenging to model in an accurate and reliable fashion.

\section{Construction of the NENCI-2021 Database \label{sec:construction_of_nenci}}

\subsection{Selection of Molecular Dimers \label{subsec:selection_of_complexes}}

\begin{figure*}[ht]
    \centering
    \includegraphics[width=\textwidth]{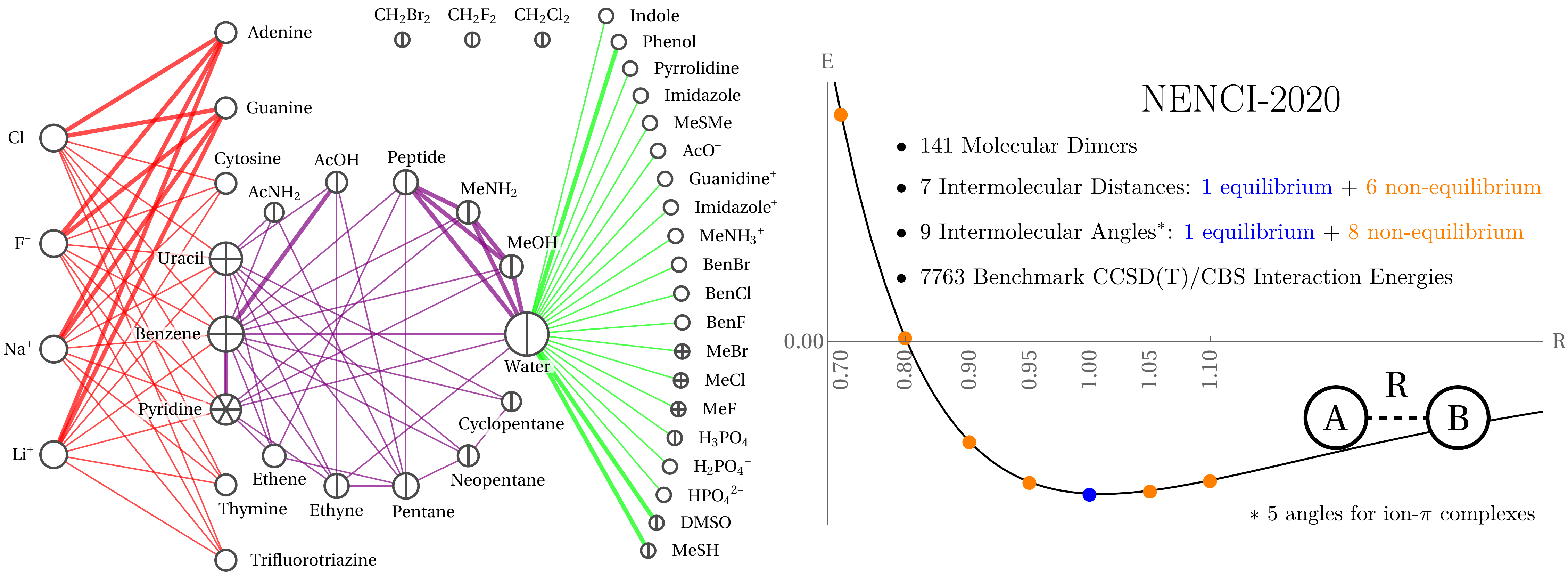}
    \caption{
        (\textit{Left}) Graphical depiction of the $141$ molecular dimers in the \NENCI database.
        \NENCI contains all of the molecular dimers in the original S66 database~\cite{rezac_s66_2011} (purple lines), the additional dimers present in the S101 (superset) database~\cite{wang_s101x7_2015} (green lines, S66 $\subset$ S101), as well as a new set of $40$ cation- and anion-$\pi$ complexes (red lines, S66 $\subset$ S101 $\subset$ \NENCI).
        In this graph, each monomer is represented by a vertex, the size of which is proportional to the number of molecular dimers involving that monomer; graph edges connecting two vertices indicate a molecular dimer formed from the connected monomers.
        Bold edges between vertices denote two different molecular dimer orientations involving the connected monomers (\eg for water--phenol, water is the hydrogen-bond donor in one dimer and the hydrogen-bond acceptor in the other). 
        Chords passing through the center of a vertex indicate a molecular dimer formed from a single monomer (\eg there is one water dimer, two uracil dimers, and three pyridine dimers in \NENCI).
        (\textit{Right}) Overall description of the \NENCI database.
        For each of the $141$ dimers described above, \NENCI generates a series of equilibrium and non-equilibrium configurations by \textit{simultaneously} sampling seven intermolecular distances and nine intermolecular angles (five for the ion-$\pi$ complexes due to symmetry considerations).
        As such, \NENCI includes ${7,763}$ benchmark CCSD(T)/CBS intermolecular interaction energies, which correspond to a wide range of equilibrium and non-equilibrium (both repulsive and attractive) geometries and emphasize close intermolecular contacts.
        See Secs.~\ref{subsec:selection_of_complexes}--\ref{sec:comp_details} for the details regarding the construction of \NENCI.
    }
    \label{fig:database}
\end{figure*}
\NENCI is a large database of $\approx {8,000}$ benchmark intermolecular interaction energies ($E_{\rm int}$, see Sec.~\ref{sec:comp_details}) that includes a diverse selection of molecular dimers and binding motifs of biological and chemical relevance, with an emphasis on non-equilibrium (attractive and repulsive) configurations and close intermolecular contacts.
As depicted in the left panel of Fig.~\ref{fig:database}, the construction of \NENCI starts with the $101$ molecular dimers in the S101~\cite{wang_s101x7_2015} database (a superset containing the earlier constructed S66~\cite{rezac_s66_2011} database), which were carefully chosen to contain small molecules with the NCIs found in biological and chemical systems.
As such, \NENCI inherits the extensive sampling of molecule types in S66 and S101, which are comprised of the \{H, C, N, O, F, P, S, Cl, Br\} atom types, range in size from small (\eg \ce{H2O}, ethene, ethyne) to medium (\eg uracil, indole, pentane), and include second- and third-row elements (\eg DMSO, MeCl, BenBr) as well as positively- (\eg \ce{MeNH3+}, \ce{Imidazole+}, \ce{Guanidine+}) and negatively- (\eg \ce{AcO-}, \ce{H2PO4-}, \ce{HPO4^2-}) charged species.
In addition, \NENCI also inherits a wide variety of intermolecular binding motifs, including dimers with single and double hydrogen bonds, halogen bonds,~\cite{Politzer_halogen_bonds_2010} and X$\mathrm{-}$H$\cdots\pi$ interactions, as well as intermolecular complexes primarily bound by dispersion, electrostatics (\eg ion-dipole, dipole-dipole, etc), and mixtures thereof.
Another salient benefit of using S66 and S101 as the foundation for \NENCI is the large number of dimers involving water, which provides a crucial first step towards the generation of benchmark intermolecular interaction energies in an aqueous environment.

\NENCI extends these databases in the following two ways: (\textit{i}) it includes $40$ new cation- and anion-$\pi$ complexes for a total of $141$ molecular dimers, and (\textit{ii}) it systematically samples both equilibrium and non-equilibrium intermolecular distances as well as intermolecular angles for each dimer (with a particular emphasis on close intermolecular contacts) for a total of ${7,763}$ benchmark interaction energies.
In particular, \NENCI includes ion-$\pi$ complexes comprised of the simplest biologically relevant monovalent cations (\Li, \Na) and anions (\F, \Cl) interacting with a representative set of $\pi$-systems, which includes the five DNA/RNA nucleobases (adenine, cytosine, guanine, thymine, uracil) as well as benzene, pyridine, and trifluorotriazine.
The inclusion of ion-$\pi$ complexes in \NENCI was primarily driven by the fact that ion-$\pi$ interactions are among the strongest NCIs known (with intermolecular interaction energies often rivaling that of hydrogen bonds and salt bridges) and have been observed throughout chemistry and biology.~\cite{Dougherty_in_bio_chem_1996,Dougherty_Cation-pi_Review_1997,Mahadevi_cation-pi_Review_2012,Anion_Pi_Perspective,Anion_Pi_Review}
This extension was also motivated by some of our recent work,~\cite{DNA_IonPi_2020} which used SAPT~\cite{SAPT_Review_Jeziorski_1994} to demonstrate that cation-$\pi$ complexes are primarily bound by induction, while anion-$\pi$ complexes are bound by a complex interplay between induction, dispersion, and electrostatics; as such, their inclusion substantially expands the scope/range of intermolecular binding motifs in \NENCI (see Sec.~\ref{subsec:SAPT}).
As shown in \pII~\cite{NENCI_part_II} of this series, this complex interplay between intermolecular forces (in addition to the presence of charged atomic species) in ion-$\pi$ complexes poses a unique challenge when trying to obtain accurate and reliable intermolecular interaction energies using both WFT and DFT methods.
In addition, the inclusion of promiscuous ion-$\pi$ binders (\ie $\pi$-systems such as the DNA/RNA nucleobases, which can form favorable ion-$\pi$ complexes with both cations and anions~\cite{DNA_IonPi_2020}) as well as $\pi$-systems that can only form energetically favorable ion-$\pi$ complexes with cations (\eg benzene) or anions (\eg trifluorotriazine) is also well-aligned with one of the fundamental goals of \NENCI: to provide a more comprehensive sampling of both attractive and repulsive non-equilibrium configurations containing a diverse array of NCI types.

Motivated by the S22x5,~\cite{Hobza_S22x5_Method_Comparison_2010} S66x8,~\cite{rezac_s66_2011,JMartin_S66x8_2016} and S101x7~\cite{wang_s101x7_2015} databases, in which intermolecular interaction energy curves were constructed for each molecular dimer, as well as the S66a8~\cite{rezac_s66a8_2011} database, in which the intermolecular angles were sampled, \NENCI systematically samples both equilibrium and non-equilibrium intermolecular distances as well as intermolecular angles for each of the $141$ molecular dimers described above.
As depicted in the right panel of Fig.~\ref{fig:database}, \NENCI samples seven intermolecular distances (\ie $0.7\times, 0.8\times, 0.9\times, 0.95\times, 1.0\times, 1.05\times, 1.1\times$ the equilibrium intermolecular separation) and nine intermolecular angles (only five intermolecular angles for the ion-$\pi$ complexes, \textit{vide infra}); for more details, see Sec.~\ref{subsec:geometries}.
\NENCI therefore contains benchmark intermolecular interaction energies (see Secs.~\ref{sec:comp_details} and \ref{sec:benchmarkerror}) for $7 \times 9 = 63$ geometries (configurations) for each of the $101$ molecular dimers in the S101 database, and $7 \times 5 = 35$ geometries for each of the $40$ ion-$\pi$ complexes, yielding a total of $63 \times 101 + 35 \times 40 = {7,763}$ equilibrium and non-equilibrium intermolecular complexes.
By including such a systematic sampling of equilibrium and non-equilibrium structures, \NENCI is a relatively large database that contains a wide range of attractive and repulsive intermolecular interaction energies (see Sec.~\ref{sec:energies_and_contacts}); as such, we believe that \NENCI will be well-suited for in-depth studies of the NCIs found throughout biology and chemistry, as well as training and testing next-generation density functional approximations, dispersion corrections, polarizable force fields, and ML-based potentials.
By including an extensive set of angularly sampled geometries at $0.7\times$ and $0.8\times$ the equilibrium intermolecular separation, \NENCI also includes a wide range of close intermolecular contacts, which are found throughout chemistry and chemical biology, as well as high-pressure systems; here, we stress again that benchmark intermolecular interaction energies in this regime not only serve as surrogates for larger/more complex systems at equilibrium, but are also important to ensure similar performance across the entire intermolecular PES when training, testing, and developing novel theoretical methods.

\subsection{Generation of Equilibrium and Non-Equilibrium Intermolecular Complexes \label{subsec:geometries}}

\begin{figure*}[ht]
    \centering
    \includegraphics[width=\textwidth]{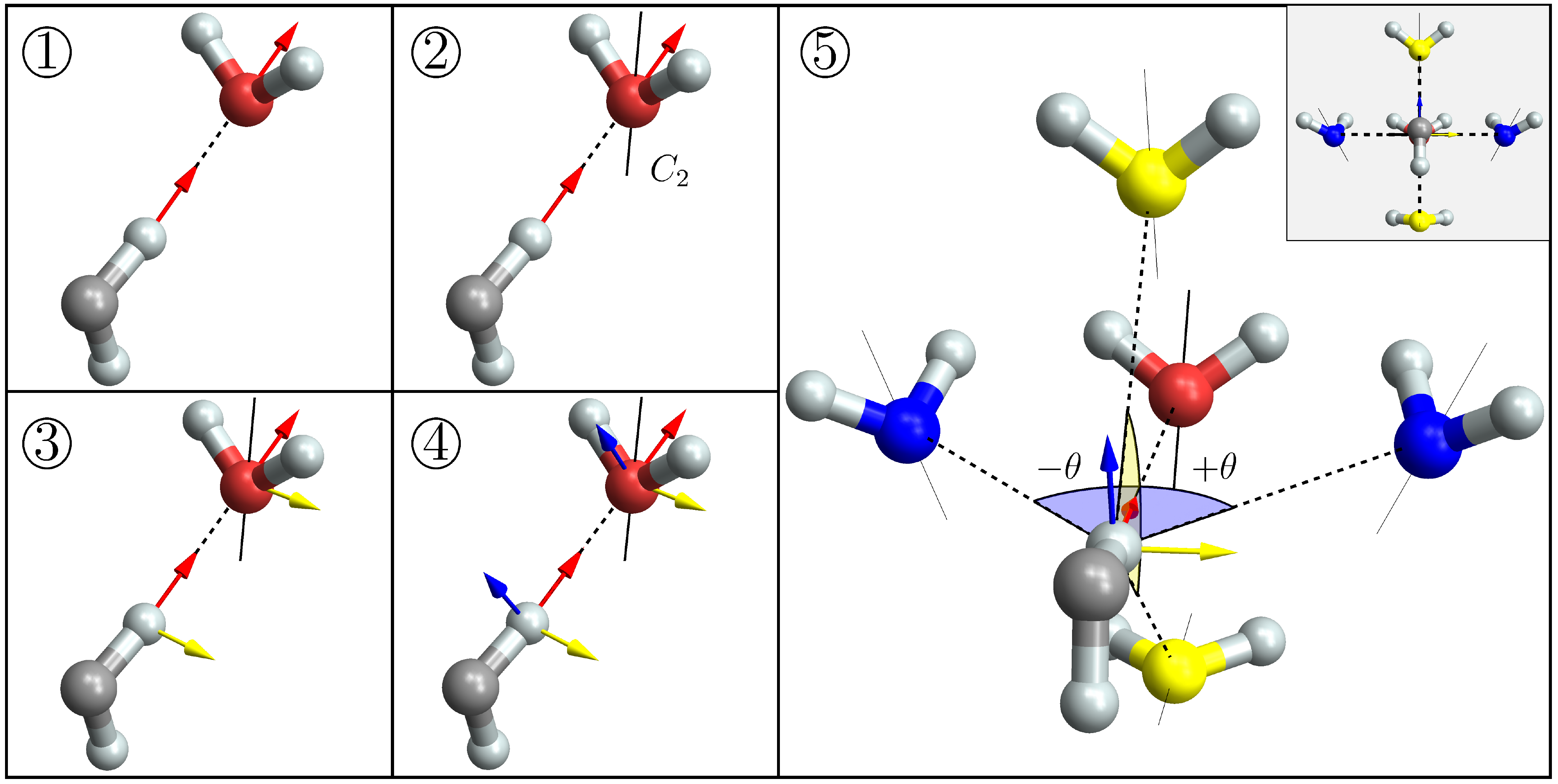}
    \caption{
        Graphical depiction of STEPS 1--5 in the protocol used for generating four (of eight) non-equilibrium intermolecular angles for the water dimer.
        As described in the text, a local reference frame ($x$-axis: solid blue arrow; $y$-axis: solid yellow arrow; $z$-axis: solid red arrow) is defined with respect to the characteristic intermolecular interaction vector (dashed black line) between monomers $A$ (red) and $B$ (gray), as well as the principal axis on monomer $A$ (solid black line).
        Preliminary geometries for the first four non-equilibrium intermolecular angles are then obtained by rotating $A$ around the $x$- and $y$-axes on $B$ by $\theta = \pm \, 30^{\circ}$. 
        For clarity, the inset to STEP 5 also provides a view down the $z$-axis of the corresponding non-equilibrium geometries.
        To obtain preliminary geometries for the remaining four non-equilibrium intermolecular angles, this procedure is repeated after swapping the monomer labels.
        See Sec.~\ref{subsec:geometries} for more details.
    }
    \label{fig:rotFig}
\end{figure*}
Unless otherwise specified, all monomer geometries were taken from the S66~\cite{rezac_s66_2011} and S101~\cite{wang_s101x7_2015} databases. 
For the eight $\pi$-systems used to construct the $40$ ion-$\pi$ complexes in \NENCI, the monomer geometries for benzene, pyridine, and uracil were taken from the S66~\cite{rezac_s66_2011} database, while the monomer geometries for the DNA/RNA nucleobases were taken from our recent work on promiscuous ion-$\pi$ binding.~\cite{DNA_IonPi_2020} 
During the construction of the equilibrium and non-equilibrium molecular dimer geometries, we employed the frozen monomer convention in which all monomers were kept fixed at their optimized geometries.

The $66$ molecular dimer geometries in the S66~\cite{rezac_s66_2011} database were also taken as is and without any changes; for the remaining $75$ molecular dimers, equilibrium geometries were optimized (see Sec.~\ref{sec:comp_details}) along a pre-defined characteristic intermolecular interaction vector.
This characteristic intermolecular interaction vector was based on the interaction type (\eg hydrogen-bonded, halogen-bonded, dispersion-bound, ion-$\pi$, etc) assigned to the molecular dimer \via chemical intuition.
Dimers that appear in the S66 database were assigned the same interaction type as in the original work,~\cite{rezac_s66_2011,rezac_s66a8_2011} and the remaining dimers were assigned an interaction type that was as consistent as possible with the S66 convention.
Given one of the following interaction types, the characteristic intermolecular interaction vector was defined as:
\begin{itemize}
    \item For hydrogen- (halogen-) bonded systems, the interaction vector points between the hydrogen (halogen) bond donor and the hydrogen (halogen) bond acceptor. For double-hydrogen-bonded systems, the interaction vector is defined as the mean of the two hydrogen-bond vectors (with both taken to originate from the same monomer).
    \item For dispersion-bound systems, the interaction vector points from the center of mass of monomer $A$ to the center of mass of monomer $B$.
    \item For ion-$\pi$ complexes, the interaction vector points from the ion to the nuclear center of charge of the $\pi$-system (computed using only the atoms in each ring, \ie the five carbons and nitrogen in pyridine). 
    Here, we note in passing that this on-axis placement of the ion does not necessarily correspond to the lowest energy geometry of each ion-$\pi$ complex.~\cite{Sherrill_Off_Center_Cation-pi,Novotny_LP-pi_analysis_and_bio_2016}
    \item Finally, there remain a few special cases (\ie the T-shaped benzene dimer), which do not fit well into any of these categories. Such systems are treated analogously with the dispersion-bound complexes, but only a subset of atoms is used in calculating an effective ``molecular center'' to ensure that the interaction vector accurately characterizes the interaction. For reference, the atoms used to calculate the interaction vector for each such complex are provided in Table~S1.
\end{itemize}
All remaining equilibrium molecular dimer geometries were obtained by minimizing the intermolecular interaction energy by rigidly translating monomer $A$ along the characteristic intermolecular interaction vector (see Sec.~\ref{sec:comp_details}), and then used as starting points to generate all non-equilibrium structures.

To systematically sample both intermolecular distances and intermolecular angles for the $141$ molecular dimers in \NENCI, we started with the procedure devised by \v{R}ez\'{a}\v{c}, Riley, and Hobza when constructing the S66x8~\cite{rezac_s66_2011} and S66a8~\cite{rezac_s66a8_2011} databases, and extended this protocol to accommodate a broader range of intermolecular interaction types and orientations.
As such, the $528$ molecular dimer geometries in the S66a8~\cite{rezac_s66a8_2011} database were also taken as is and without any changes.
The procedure for generating the remaining ${7,094}$ non-equilibrium intermolecular complexes in \NENCI is outlined below, with STEPS 1--5 graphically illustrated for the water dimer in Fig.~\ref{fig:rotFig}: \\[-0.5em]

\noindent \textbf{STEP 1.} Starting with an optimized equilibrium intermolecular complex, arbitrarily label each monomer as either $A$ or $B$ (except for the ion-$\pi$ complexes, in which the ion should be labelled as monomer $A$).
Draw the characteristic intermolecular interaction vector from $B$ to $A$ (dashed black line) according to the interaction type assigned to the molecular dimer (\textit{vide supra}).
Define the $z$-axis (solid red arrow) along the interaction vector. \\[-0.5em]

\noindent \textbf{STEP 2.} Without loss of generality, assume that $A$ will be rotated around $B$ (the alternative will be dealt with in STEP 8 below).
To determine the axes of rotation, first find the principal axis ($C_n$) corresponding to monomer $A$ (\ie the molecular axis with the highest degree ($n$) of rotational symmetry); for the water monomer depicted in Fig.~\ref{fig:rotFig}, the principal axis is the solid black line labelled $C_2$.
If no principal axis with $n \ge 2$ exists, we follow the convention used during the construction of the S66a8~\cite{rezac_s66a8_2011} database, \ie an approximate principal axis is defined by removing all hydrogen atoms from the molecule and reducing the identity of each heavy atom and functional group to identical spheres. \\[-0.5em]

\noindent \textbf{STEP 3.} Define the $y$-axis (solid yellow arrow) to be perpendicular to the $z$-axis and the principal axis of $A$. \\[-0.5em]

\noindent \textbf{STEP 4.} Define the $x$-axis (solid blue arrow) to be perpendicular to the $z$- and $y$-axes, thereby completely specifying the local reference frame used in this work. \\[-0.5em]

\noindent \textbf{STEP 5.} To generate preliminary geometries for the first four non-equilibrium intermolecular angles, rotate $A$ about the $x$- and $y$-axes passing through the tail of the interaction vector (\ie located on monomer $B$) by $\theta = \pm \, 30^\circ$. \\[-0.5em]

\noindent \textbf{STEP 6.} For each non-equilibrium intermolecular angle, minimize the intermolecular interaction energy by rigidly translating $A$ along the characteristic intermolecular interaction vector (see Sec.~\ref{sec:comp_details}).
For the ion-$\pi$ complexes that are repulsive along the entire dissociation curve (\eg \ce{Na+}$\cdots$trifluorotriazine), the minimum of the SAPT exchange + induction + dispersion (EID) energy~\cite{DNA_IonPi_2020} was used in lieu of the intermolecular interaction energy (see Sec.~\ref{sec:comp_details}).
Define the intermolecular distance (\ie the length of the characteristic intermolecular interaction vector) in each optimized geometry as the equilibrium ($1.0\times$) intermolecular distance for the given non-equilibrium intermolecular angle. \\[-0.5em]

\noindent \textbf{STEP 7.} For each non-equilibrium intermolecular angle, scale the corresponding (optimized) interaction vector by factors of $0.7\times$, $0.8\times$, $0.9\times$, $0.95\times$, $1.05\times$, and $1.1\times$, and rigidly translate $A$ consistent with each scaled vector.
This will provide molecular dimer geometries along four separate intermolecular dissociation curves corresponding to each of the four non-equilibrium intermolecular angles. \\[-0.5em]

\noindent \textbf{STEP 8.} Switch the $A$ and $B$ labels, and repeat STEPS 1--7.
This will provide molecular dimer geometries along the intermolecular dissociation curves corresponding to each of the remaining four non-equilibrium intermolecular angles (for a total of eight non-equilibrium intermolecular angles). 
N.B.: The ion-$\pi$ complexes in \NENCI only have four unique non-equilibrium intermolecular angles due to the spherical symmetry of the ion; as such, STEP 8 is unnecessary and can be skipped for these molecular dimers. \\[-0.5em]

\noindent \textbf{STEP 9.} For the equilibrium intermolecular angle, also scale the corresponding (optimized) interaction vector by factors of $0.7\times$, $0.8\times$, $0.9\times$, $0.95\times$, $1.05\times$, and $1.1\times$, and rigidly translate $A$ consistent with each scaled vector.
This will provide molecular dimer geometries along the intermolecular dissociation curve corresponding to the equilibrium intermolecular angle. \\[-0.5em]

\subsection{Computational Details \label{sec:comp_details}}

Intermolecular interaction energies ($E_{\rm int}$) for each of the ${7,763}$ intermolecular complexes in \NENCI were computed \via
\begin{align}
    E_{\rm int} &= E_{AB} - E_{A} - E_{B} ,
    \label{eq:Eint}
\end{align}
in which $E_{AB}$ is the total energy of the dimer and $E_A$ ($E_B$) is the total energy of monomer $A$ ($B$).
As mentioned above, all monomers were kept fixed at their optimized geometries, and the counterpoise correction of Boys and Bernardi~\cite{Boys-Bernardi_CP} was applied to minimize basis set superposition error (BSSE).

Unless otherwise specified, Dunning's correlation consistent basis sets (with and without diffuse functions), namely cc-pVXZ and aug-cc-pVXZ (with X = D, T, Q),~\cite{basis_Dunning_aXZ_H_Ne_1989,Dunning_Diffuse_aXZ_1992,basis_Dunning_aXZ_Al_Ar_1993,Dunning_aXZ_basis_Ga_Kr_1999} along with the frozen core (FC) approximation were used for all atoms except Li and Na.
To provide a more accurate description of the core/valence electrons in the cation-$\pi$ complexes, the cc-pwCVXZ~\cite{basis_pwcvtz_Li_Na_2011} and aug-cc-pwCVXZ basis sets~\cite{basis_pwcvtz_Li_Na_2011} were used for Li and Na in conjunction with the following modified FC approximation: Li$^+$ = 1s$^2$ (no core) and Na$^+$ = [He]2s$^2$2p$^6$ ([He] core).
All calculations employed the resolution-of-the-identity (RI) or density-fitting (DF) approximation during self-consistent field (SCF) calculations at the mean-field Hartree-Fock (HF) level as well as during post-HF calculations to account for electron correlation effects; the RI/DF approximation has been shown to introduce negligible errors when computing intermolecular interaction energies.~\cite{Hobza_RI-MP2_Accuracy_DNA_Stacks_2001,Sherrill_DF_CCSD(T)_Accuracy_2013}
Whenever available, the corresponding JKFIT and RI auxiliary basis sets were used in conjunction with each primary (atomic orbital) basis set, \ie cc-pVXZ-JKFIT/cc-pVXZ-RI~\cite{basis_Weigend_aXZ_JK_2002,basis_Weigend_aXZ_RI_2002} were used with cc-pVXZ, and aug-cc-pVXZ-JKFIT/aug-cc-pVXZ-RI~\cite{basis_Weigend_aXZ_JK_2002,basis_Weigend_aXZ_RI_2002} were used with aug-cc-pVXZ.
For the cation-$\pi$ complexes, the def2-aQZVPP-JKFIT/def2-aQZVPP-RI auxiliary basis sets~\cite{basis_Hattig_RI_2005,basis_Weigend_aux_jk_basis_2008,Hellweg_development_qzvpp-ri_2015} (which are some of the largest available auxiliary basis sets) were taken from the MOLPRO~\cite{MOLPRO-WIREs,MOLPRO} basis set library and used in conjunction with cc-pwCVXZ and aug-cc-pwCVXZ for Li and Na.
Throughout this work, we used the abbreviation aXZ to denote the following basis set usage: aug-cc-pVXZ (with aug-cc-pVXZ-JKFIT/aug-cc-pVXZ-RI) for \{H, C, N, O, F, S, P, Cl, Br\}; aug-cc-pwCVXZ (with def2-aQZVPP-JKFIT/def2-aQZVPP-RI) for \{Li, Na\}; we also use the abbreviation haXZ (\ie heavy-aug-cc-pVXZ, also known as jul-cc-pVXZ~\cite{Truhlar_calendar_basis_sets_2011}) to mean: cc-pVXZ (with cc-pVXZ-JKFIT/cc-pVXZ-RI) for \{H\}, aug-cc-pVXZ (with aug-cc-pVXZ-JKFIT/aug-cc-pVXZ-RI) for \{C, N, O, F, S, P, Cl, Br\}, and aug-cc-pwCVXZ (with def2-aQZVPP-JKFIT/def2-aQZVPP-RI) for \{Li, Na\}.

For each molecular dimer and non-equilibrium intermolecular angle , the corresponding optimal intermolecular distance ($1.0\times$) was obtained \via a constrained minimization of $E_{\rm int}$ at the BSSE-corrected MP2/cc-pVTZ level (see Eq.~\eqref{eq:Eint}); for the molecular dimers not included in the original S66 database, the same procedure was also used to obtain the optimal intermolecular distance for the equilibrium intermolecular angle. 
In practice, this was accomplished by computing $E_{\rm int}$ for a series of dimer geometries in which monomer $A$ (and/or $B$) was rigidly translated along the characteristic intermolecular interaction vector (see Sec.~\ref{subsec:geometries}), and then locating the minimum value along the corresponding cubic spline interpolant.

Benchmark $E_{\rm int}$ values in \NENCI were obtained using Eq.~\eqref{eq:Eint} with all dimer ($E_{AB}$) and monomer ($E_A$ and $E_B$) contributions computed using the ``gold standard'' CCSD(T) method extrapolated to the complete basis set (CBS) limit,~\cite{Hobza_ccsdt_gold_standard_2013,Sherrill_DeltaCC_extrapolation_2011} \ie
\begin{align}
    E^{\rm CCSD(T)/CBS} \equiv E^{\rm MP2/CBS} + \delta E^{\rm CCSD(T)/haTZ} . 
    \label{eqn:CBSExtrap1}
\end{align}
In this expression, the CBS-extrapolated MP2 total energy,  
\begin{align}
    E^{\rm MP2/CBS} &\equiv E^{\rm MP2/a(TQ)Z} \nonumber \\
    &= E^{\rm HF/aQZ} + E^{\rm MP2/a(TQ)Z}_{\rm corr} ,
    \label{eqn:CBSExtrap2}
\end{align}
was obtained using the two-point extrapolation procedure of Halkier \textit{et al.}~\cite{Helgaker_Basis_Set_Convergence_1998} on the MP2 correlation energy, namely,
\begin{align}
    E^{\rm MP2/a(XY)Z}_{\rm corr} = \frac{X^3 E_{\rm corr}^{\rm MP2/aXZ} - Y^3 E_{\rm corr}^{\rm MP2/aYZ}}{X^3-Y^3}
    \label{eqn:CBSExtrap3}
\end{align}
with $X = 3$ (aTZ) and $Y = 4$ (aQZ). The so-called ``delta'' CCSD(T) correction, 
\begin{align}
    \delta E^{\rm CCSD(T)/haTZ} = E^{\rm CCSD(T)/haTZ} - E^{\rm MP2/haTZ} ,
    \label{eqn:CBSExtrap4}
\end{align}
was computed using the haTZ basis set.
The accuracy of this scheme for computing $E_{\rm int}$---in particular for intermolecular complexes with particularly close contacts (\ie $0.7\times$ the equilibrium intermolecular separation)---is critically assessed below in Sec.~\ref{sec:benchmarkerror}.

The energy decomposition analysis scheme (and classification of intermolecular binding motifs) provided in Sec.~\ref{subsec:SAPT} was based on calculations at the SAPT2+/aDZ level of theory,~\cite{SAPT_Review_Jeziorski_1994,SAPT0_2010,SAPT0_large_2011,Density_Fitting_SAPT_Sherrill_2010,SAPT_Nat_orbs_2010} the so called ``silver standard'' of SAPT.~\cite{Sherrill_SAPT_Levels_2014}

All calculations in this work were performed using the \texttt{Psi4} (\texttt{v1.2}) software program.~\cite{Psi4_11_2017}
During all HF calculations, the SCF convergence parameters were set to $1.0 \times 10^{-8}$ in the total energy (\texttt{e\_convergence = 1E-8}) and $1.0 \times 10^{-8}$ in the root-mean-square DIIS error (\texttt{d\_convergence = 1E-8}).
For all CCSD(T) calculations, the CCSD convergence parameters were set to $1.0 \times 10^{-6}$ in the total energy (\texttt{e\_convergence = 1E-6}) and $1.0 \times 10^{-5}$ in the residual of the $t$-amplitudes (\texttt{r\_convergence = 1E-5}). 

\subsection{Obtaining the NENCI-2021 Database \label{sec:obtaining}}

A single \texttt{zip} file containing the Cartesian coordinates of the ${7,763}$ intermolecular complexes in \NENCI (in \texttt{xyz} format) is provided in the Supplementary Material. 
The properties of each monomer (\ie charge, multiplicity, number of atoms), the corresponding benchmark $E_{\rm int}$ value, as well as the CCSD(T)/CBS and SAPT energetic components can be found in the comment line of each \texttt{xyz} file (see \texttt{README} file for additional details).

\section{Properties of the NENCI-2021 Database \label{sec:properties}}

\subsection{Statistical Analysis of Intermolecular Interaction Energies and Closest Intermolecular Contacts \label{sec:energies_and_contacts}}

\begin{figure}[t]
    \centering
    \includegraphics[width=\columnwidth]{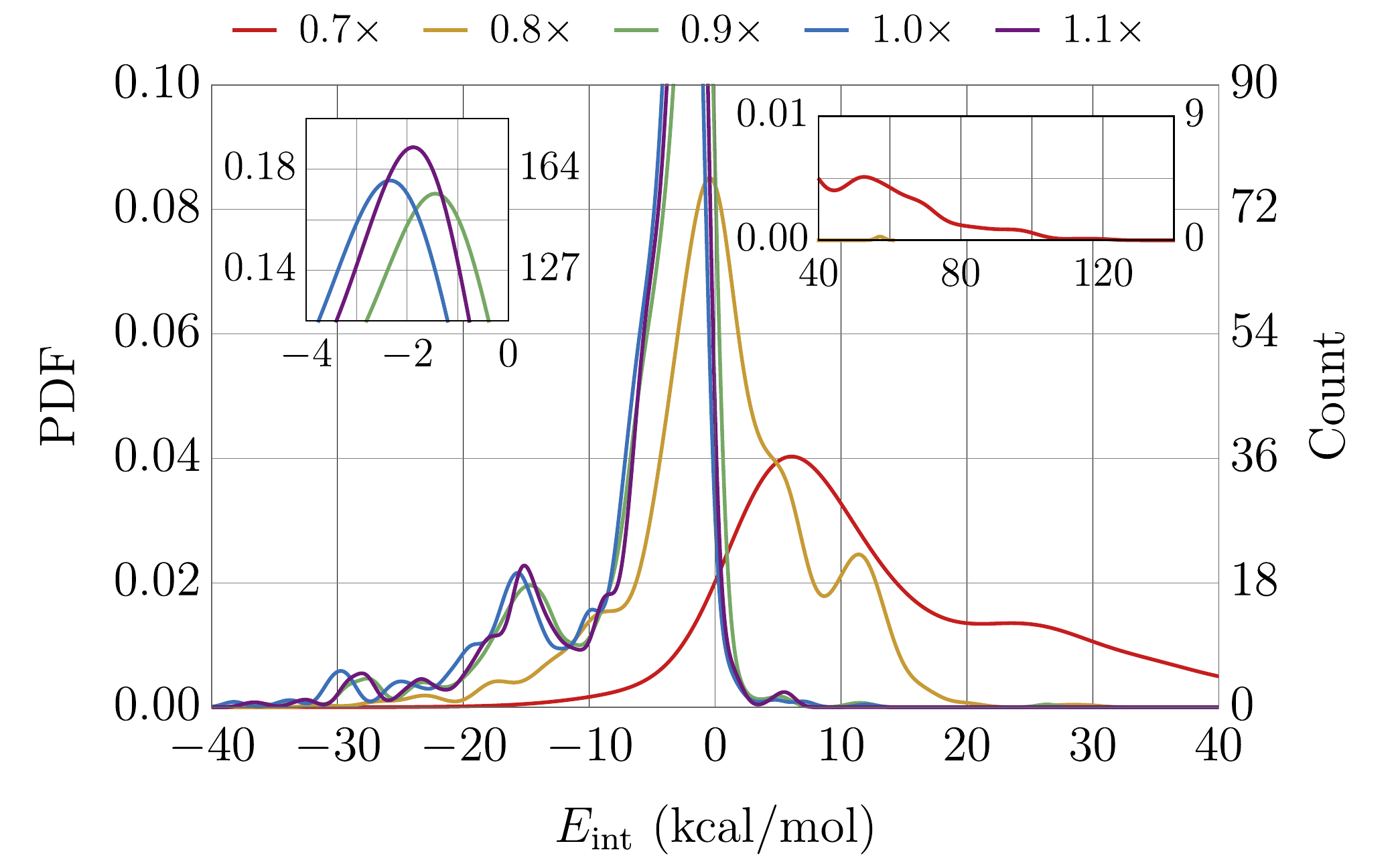}
    \caption{
    Normalized probability density functions (PDFs) of the benchmark $E_{\rm int}$ values in the \NENCI database as a function of the intermolecular distance (with the $0.95 \times$ and $1.05 \times$ scaled intermolecular distances omitted for clarity).
    The $E_{\rm int}$ values in \NENCI range from $-38.5$~kcal/mol (most attractive) to $+186.8$~kcal/mol (most repulsive), with a mean (median) interaction energy of $-1.06$~kcal/mol ($-2.39$~kcal/mol).
    Insets display the peaks of the $0.9 \times$, $1.0 \times$, and $1.1 \times$ PDFs as well as the (positive $E_{\rm int}$) tails of the $0.7 \times$ and $0.8 \times$ PDFs.
    \label{fig:EIntDistro}}
\end{figure}

A well-balanced database of intermolecular interactions should have a wide range of $E_{\rm int}$ values,~\cite{rezac_s66_2011} and this is indeed the case for \NENCI, as evidenced by the normalized $E_{\rm int}$ distributions provided in Fig.~\ref{fig:EIntDistro}. 
With $E_{\rm int}$ values ranging from $-38.5$~kcal/mol to $+186.8$~kcal/mol, the benchmark intermolecular interaction energies in \NENCI span $225.3$~kcal/mol. 
In general, the most attractive (most negative) $E_{\rm int}$ values in \NENCI correspond to charged intermolecular complexes that tend to be at (or close) to their equilibrium geometries.
For instance, the single most attractive intermolecular complex in \NENCI (with $E_{\rm int} = -38.5$~kcal/mol) corresponds to the \ce{Li+}$\cdots$benzene ion-$\pi$ system at its equilibrium geometry (\ie with \ce{Li+} located above the center of the benzene ring, see Sec.~\ref{subsec:geometries}). 
In fact, the top ten most attractive intermolecular interactions in \NENCI correspond to the various \ce{Li+}$\cdots\pi$ complexes at slightly different (but close to equilibrium) intermolecular distances and angles; these are followed by the ionic \ce{H2O}$\cdots$\ce{HPO4^2-} hydrogen-bonded complexes (with a minimum $E_{\rm int} = -34.6$~kcal/mol) and the \ce{Na+}$\cdots\pi$ complexes (with a minimum $E_{\rm int} = -28.2$~kcal/mol).
In general, the most repulsive (most positive) $E_{\rm int}$ values in \NENCI correspond to intermolecular complexes in which the monomers are separated by the shortest distance ($0.7 \times$) and rotated away from their equilibrium intermolecular angle, as both of these geometric perturbations lead to a rapid increase in the exponentially repulsive steric contribution to the interaction energy.
For instance, the single most repulsive intermolecular complex in \NENCI (with $E_{\rm int} = +186.8$~kcal/mol) corresponds to the dimethyl sulfoxide (DMSO) dimer separated by $0.7 \times$ the equilibrium intermolecular distance and rotated to a non-equilibrium angle; in fact, this dimer has the closest intermolecular contact in the entire database (with $d_{\rm H \cdots H} = 0.81$~\AA, \textit{vide infra}).
Other substantially repulsive intermolecular complexes in \NENCI include the \ce{Na+}$\cdots$trifluorotriazine ion-$\pi$ system (with a maximum $E_{\rm int} = +118.8$~kcal/mol) and another DMSO dimer (with $E_{\rm int} = +112.4$~kcal/mol), both of which were characterized by a $0.7 \times$ intermolecular separation and a non-equilibrium intermolecular angle.

The mean and median $E_{\rm int}$ values in \NENCI are $-1.1$~kcal/mol and $-2.4$~kcal/mol, respectively, which correspond to typical interaction energies found in weakly bound molecular dimers.
These statistical measures are primarily governed by the (relatively) large number of intermolecular complexes in \NENCI that contain monomers in non-equilibrium (angular) orientations.
Such geometric perturbations tend to nullify the energetic stabilization provided by \textit{directional} intermolecular binding motifs (\eg single- and double-hydrogen bonds, dipole-dipole interactions, etc), and often result in complexes with weakly attractive $E_{\rm int}$ values.
Broken down by the scaled intermolecular distance, the mean (median) $E_{\rm int}$ values are: $+21.8$ ($+11.7$)~kcal/mol for $0.7 \times$, $+0.5$ ($+0.2$)~kcal/mol for $0.8 \times$, $-5.2$ ($-2.9$)~kcal/mol for $0.9 \times$, $-6.0$ ($-3.5$)~kcal/mol for $0.95 \times$, $-6.1$ ($-3.5$)~kcal/mol for $1.0 \times$, $-5.9$ ($-3.4$)~kcal/mol for $1.05 \times$, and $-5.5$ ($-3.1$)~kcal/mol for $1.10 \times$.
In total, \NENCI contains ${6,020}$ attractive ($E_{\rm int} < 0$) and ${1,743}$ repulsive ($E_{\rm int} > 0$) intermolecular complexes, and the crossover from attractive to repulsive $E_{\rm int}$ values typically occurs around $0.8 \times$ the equilibrium intermolecular distance.
As one might expect, the proportion of attractive intermolecular interactions in \NENCI quickly diminishes as the distance between monomers decreases; broken down again by the scaled intermolecular distance, we find the percentage of attractive (repulsive) $E_{\rm int}$ values are: $3.4\%$ ($96.6\%$) for $0.7 \times$, $47.8\%$ ($52.2\%$) for $0.8 \times$, $97.6\%$ ($2.4\%$) for $0.9 \times$, $98.3\%$ ($1.7\%$) for $0.95 \times$, $98.3\%$ ($1.7\%$) for $1.0 \times$, $98.2\%$ ($1.8\%$) for $1.05 \times$, and $98.0\%$ ($2.0\%$) for $1.1 \times$.
Quite interestingly, there are still a number ($N = 38$) of attractive intermolecular complexes at the $0.70 \times$ scaled intermolecular distance, which generally correspond to strongly favorable dimers such as the \ce{Li+}$\cdots\pi$ complexes discussed above.
In the same breath, there are also quite a few ($N = 19$) repulsive complexes at the equilibrium ($1.0 \times$) distance---some of which even occur at the corresponding equilibrium angle, \eg the cation- and anion-$\pi$ complexes involving trifluorotriazine and benzene, respectively.

\begin{figure}[t!]
    \centering
    \includegraphics[width=\columnwidth]{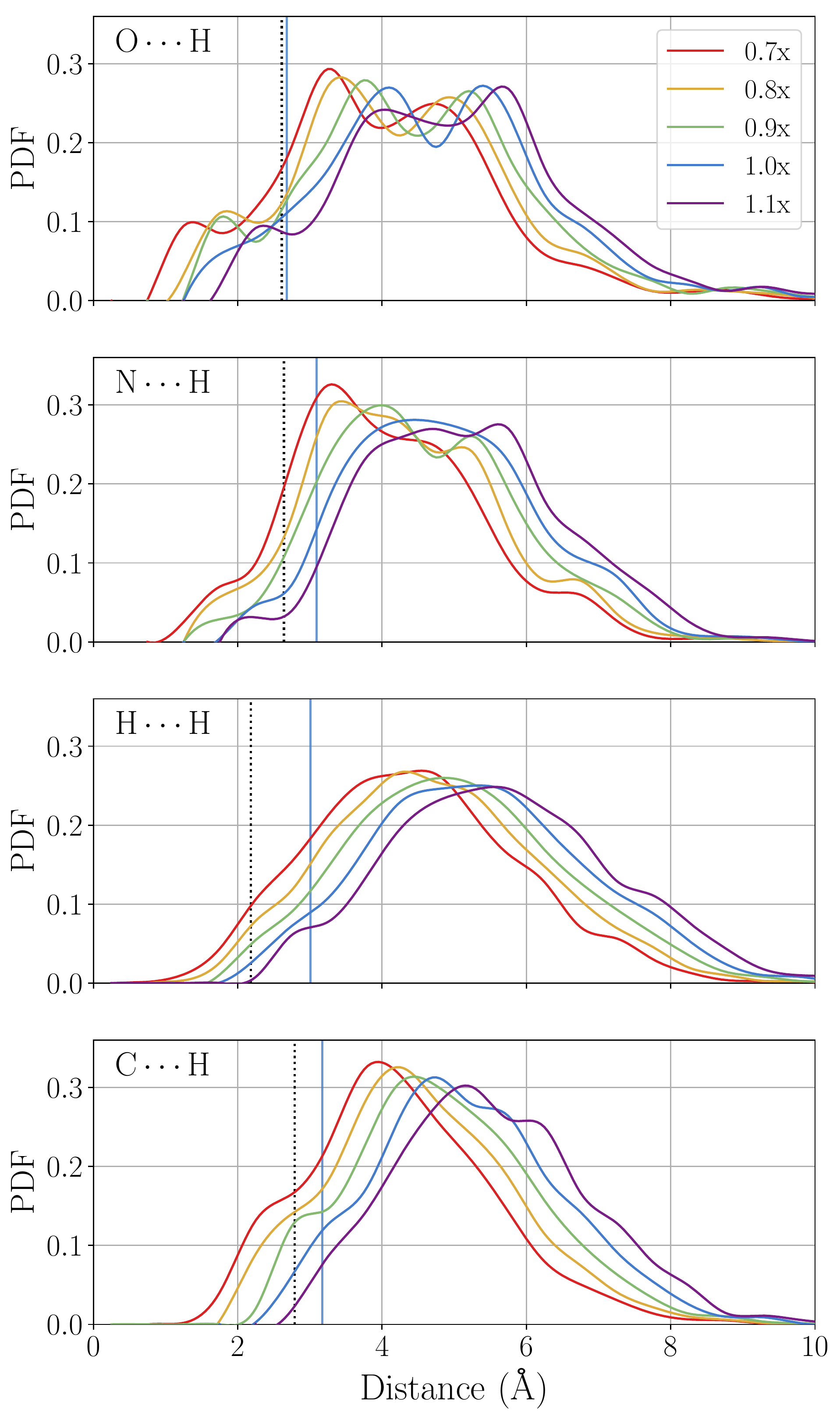}
    \caption{
    Normalized probability density functions (PDFs) of the $\rm O \cdots H$, $\rm N \cdots H$, $\rm H \cdots H$, and $\rm C \cdots H$ intermolecular atom-pair distances in the \NENCI database as a function of the scaled intermolecular distance (with the $0.95 \times$ and $1.05 \times$ scaled intermolecular distances omitted for clarity).
    For reference, vertical solid blue lines indicate the mean closest intermolecular contact distances in the $141$ equilibrium complexes in \NENCI, while the vertical dotted black lines indicate the sum of the atomic vdW radii~\cite{bondi_vdw_radii_1964,updated_vdw_radii_1996} corresponding to each atom pair.
    See main text for more details.
    }
\label{fig:rdf}
\end{figure}
A well-balanced database of intermolecular interactions should also sample a wide range of intermolecular atom-pair distances (\ie interatomic distances between the atoms on molecule $A$ and the atoms on molecule $B$).
Again, this is indeed the case for \NENCI, and is demonstrated by the series of normalized PDFs in Fig.~\ref{fig:rdf}, which quantify a representative set of atom-pair distances (\ie $\rm O \cdots H$, $\rm N \cdots H$, $\rm H \cdots H$, and $\rm C \cdots H$) as a function of intermolecular separation.
In this figure, we chose to focus on the $\rm O \cdots H$, $\rm N \cdots H$, $\rm H \cdots H$, and $\rm C \cdots H$ intermolecular atom-pair distances, as the first two are representative of hydrogen-bonded systems and the last two are the relevant interatomic distances for non-bonded complexes in general.
Since \NENCI was designed with a particular emphasis on close intermolecular contacts, we focus our discussion on the short-distance sectors in these PDFs.
As discussed above in the Introduction, such close intermolecular contacts are important in a number of applications,~\cite{structure_based_discovery_docking_2009,hobza_reliable_drug_docking_2010,med_chem_guide_to_interactions_2010,wang_s101x7_2015,Hobza_malonate_docking_w_experiment_2015} and pose significant difficulty for both WFT and DFT methods alike (see \pII~\cite{NENCI_part_II} in this series), as both strongly attractive and strongly repulsive intermolecular forces must be accurately described to obtain a quantitatively correct $E_{\rm int}$ value.
As the intermolecular distance is reduced from $1.1 \times$ to $0.7 \times$, the complexes in \NENCI sample increasingly closer interatomic distances and begin to more appreciably populate the region inside the corresponding vdW envelope.
In other words, a number of intermolecular atom-pair distances ($R_{AB}$) are less than the sum of the corresponding vdW radii, \ie $R_{AB} < R_{AB}^{\rm vdW} \equiv R_{A}^{\rm vdW} + R_{B}^{\rm vdW}$.
Plotted as vertical dotted black lines in Fig.~\ref{fig:rdf}, these $R_{AB}^{\rm vdW}$ values were computed using the vdW radii provided by Bondi~\cite{bondi_vdw_radii_1964} for \{C, N, O\} and the revised value obtained by Rowland and Taylor~\cite{updated_vdw_radii_1996} for \{H\}, and take on the following values: $2.61$~{\AA} ($\rm O \cdots H$), $2.64$~{\AA} ($\rm N \cdots H$), $2.18$~{\AA} ($\rm H \cdots H$), and $2.79$~{\AA} ($\rm C \cdots H$).
As one would expect, these values are always smaller than the mean closest contact distances for the $141$ equilibrium intermolecular complexes in \NENCI, \ie $2.68$~{\AA} ($\rm O \cdots H$), $3.09$~{\AA} ($\rm N \cdots H$), $3.01$~{\AA} ($\rm H \cdots H$), and $3.17$~{\AA} ($\rm C \cdots H$); for comparative purposes, these values are plotted as vertical solid blue lines in Fig.~\ref{fig:rdf}.
Broken down by scaled intermolecular distance, the percentage of $\rm O \cdots H$ ($\rm N \cdots H$) intermolecular atom-pair distances with $R_{AB} < R_{AB}^{\rm vdW}$ are $17.0\%$ ($10.8\%$) for $0.7 \times$, $13.5\%$ ($7.4\%$) for $0.8 \times$, $10.7\%$ ($4.4\%$) for $0.9 \times$, $9.1\%$  ($3.4\%$) for $0.95 \times$, $7.8\%$  ($2.8\%$) for $1.0 \times$, $6.9\%$  ($2.4\%$) for $1.05 \times$, and $6.1\%$  ($1.9\%$) for $1.1 \times$.
Applying the same analysis to the $\rm H \cdots H$ ($\rm C \cdots H$) atom-pair distances yields the following: $4.5\%$ ($13.2\%$) for $0.7 \times$, $2.4\%$ ($8.6\%$) for $0.8 \times$, $1.2\%$ ($4.4\%$) for $0.9 \times$, $0.7\%$ ($2.5\%$) for $0.95 \times$, $0.2\%$ ($1.2\%$) for $1.0 \times$, $0.1\%$ ($0.5\%$) for $1.05 \times$, and $0.1\%$ ($0.2\%$) for $1.1 \times$. 
The closest contacts in \NENCI occur in complexes in which the intermolecular distance has been scaled to $0.7 \times$ and the monomers have been rotated such that the atoms (on each monomer) are forced into close proximity.
For reference, the shortest intermolecular atom-pair distances in \NENCI are significantly shorter than $R_{AB}^{\rm vdW}$, and were found to be: $0.82$~\AA\ for $\rm O \cdots H$ (DMSO dimer, $E_{\rm int} = +186.8$~kcal/mol, $31\%$ of $R_{\rm OH}^{\rm vdW}$), $1.30$~\AA\ for $\rm N \cdots H$ (uracil$\cdots$neopentane, $E_{\rm int} = +80.5$~kcal/mol, $49\%$), $0.81$~\AA\ for $\rm H \cdots H$ (peptide$\cdots$pentane, $E_{\rm int} = +72.3$~kcal/mol, $37\%$), and 
$1.22$~\AA\ for $\rm C \cdots H$ (benzene$\cdots$AcOH, $E_{\rm int} = +76.4$~kcal/mol, $44\%$).
While it is not surprising to find the most repulsive intermolecular complex in \NENCI (\ie the DMSO dimer) listed among the closest contacts, the other examples are far from the most positive end of the $E_{\rm int}$ spectrum and reflect the wide range of attractive and repulsive intermolecular forces sampled in this database.

\begin{figure*}[ht!]
    \centering
    \includegraphics[width=\textwidth,trim={0cm 0cm 10cm 0cm}]{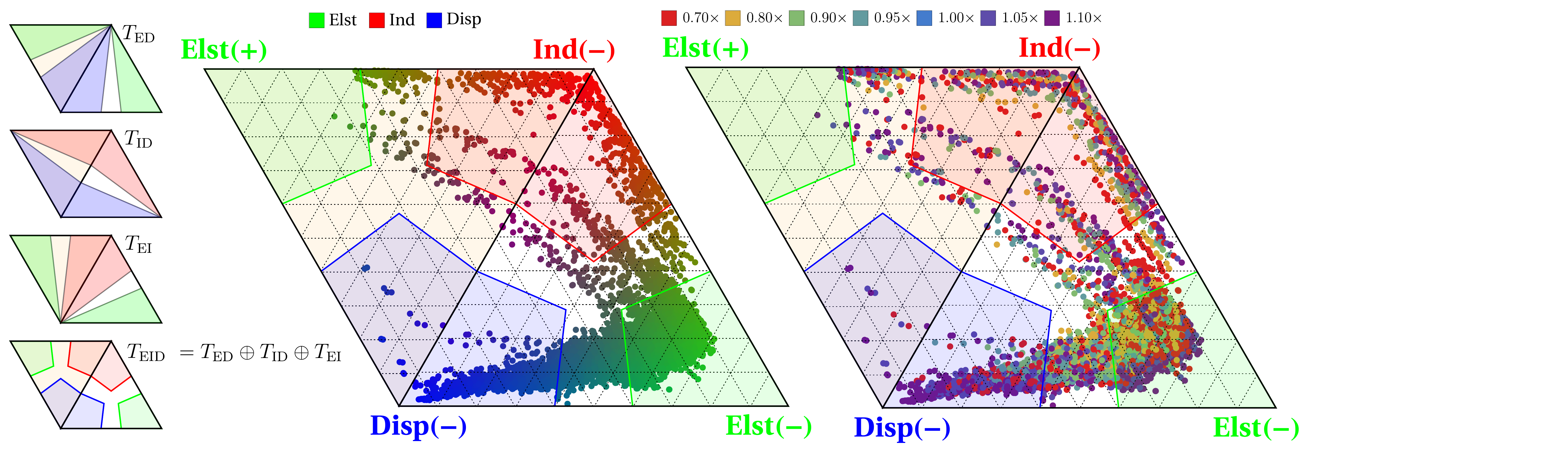}
    \caption{
    (\textit{Left}) Geometric depiction of the extended four-category classification scheme (based on a SAPT decomposition of $E_{\rm int}$) used to classify each intermolecular complex in \NENCI as: Elst-bound (E, green), Ind-bound (I, red), Disp-bound (D, blue), or Mix-bound. 
    As described in the main text, this classification scheme can be represented by a fused ternary diagram ($T_{\rm EID}$) which has been colored according to the following rule: each intermolecular complex that has been assigned the same category (color) in any two of the three sub-classification schemes ($T_{\rm ED}$, $T_{\rm ID}$, $T_{\rm EI}$) retains that color in $T_{\rm EID}$; otherwise, the complex is classified as Mix-bound (white).
    (\textit{Middle/Right}) Ternary diagrams depicting the breakdown of the SAPT2+/aDZ intermolecular interaction energies of each complex in \NENCI according to the contributions from electrostatics ($\varepsilon_{\rm Elst}$), induction ($\varepsilon_{\rm Ind}$), and dispersion ($\varepsilon_{\rm Disp}$).
    Since $\varepsilon_{\rm Elst}$ can be positive (Elst(+)) or negative (Elst(-)), these plots are comprised of two ternary diagrams (one for Elst(+) and one for Elst(-)) that have been fused together.
    In these ternary diagrams, the shaded polygons are used to reflect the four-category classification scheme described above, \ie Elst-bound (green), Ind-bound (red), and Disp-bound (blue); complexes that are not located in any one of these regions are Mix-bound.
    In the \textit{Middle} panel, each point has been colored using an RGB scheme with values given by: $\left\{ \left| \varepsilon_{\rm Ind} \right| / \left( \left| \varepsilon_{\rm Elst} \right| + \left| \varepsilon_{\rm Ind} \right| + \left| \varepsilon_{\rm Disp} \right| \right), \, \left| \varepsilon_{\rm Elst} \right| / \left( \left| \varepsilon_{\rm Elst} \right| + \left| \varepsilon_{\rm Ind} \right| + \left| \varepsilon_{\rm Disp} \right| \right), \, \left| \varepsilon_{\rm Disp} \right| / \left( \left| \varepsilon_{\rm Elst} \right| + \left| \varepsilon_{\rm Ind} \right| + \left| \varepsilon_{\rm Disp} \right| \right) \right\}$.
    In the \textit{Right} panel, each point is colored according to the scaled intermolecular distance.
    }
    \label{fig:ternary}
\end{figure*}

\subsection{Energy Decomposition Analysis of the Intermolecular Binding Motifs \label{subsec:SAPT}}

A well-balanced database of intermolecular interactions should also sample a wide variety of different binding motifs.
Here, we would again argue that this is the case for \NENCI, and demonstrate this point by the extensively populated ternary diagrams depicted in Fig.~\ref{fig:ternary}.
Introduced by Kim \textit{et al.}~\cite{Kim_pi_analysis_sapt_ternary_2009} in the late 2000s, these ternary diagrams were constructed using a SAPT decomposition of $E_{\rm int}$  into the following four components for each intermolecular complex in \NENCI: $\varepsilon_{\rm Elst}$ (electrostatics, Elst), $\varepsilon_{\rm Exch}$ (exchange, Exch), $\varepsilon_{\rm Ind}$ (induction, Ind), and $\varepsilon_{\rm Disp}$ (dispersion, Disp), \ie $E_{\rm int} \approx \varepsilon_{\rm SAPT} = \varepsilon_{\rm Elst} + \varepsilon_{\rm Exch} + \varepsilon_{\rm Ind} + \varepsilon_{\rm Disp}$.
In particular, we performed this decomposition at the SAPT2+/aDZ level of theory,~\cite{SAPT_Review_Jeziorski_1994,SAPT0_2010,SAPT0_large_2011,Density_Fitting_SAPT_Sherrill_2010,SAPT_Nat_orbs_2010} the so-called ``silver standard'' of SAPT,~\cite{Sherrill_SAPT_Levels_2014} which has been shown to have an overall mean absolute error (MAE) of $0.30$~kcal/mol across the S22,~\cite{jurecka_s22_2006} HBC6,~\cite{Sherrill_Benchmark_Hbond_2011} NBC10,~\cite{Sherrill_BZ_Dimer_2006,Sherrill_aliphatic_CH_pi_2006,Sherrill_Benchmark_disp_bound_PEC_2009,Sherrill_indole_BZ_2010} and HSG~\cite{HSG_protein_ligand_2011} databases.~\cite{Sherrill_DeltaCC_extrapolation_2011} 
Unlike the ``bronze standard'' sSAPT0/jun-cc-pVDZ,~\cite{Sherrill_SAPT_Levels_2014} which can underestimate the dispersion component in anion-$\pi$ complexes by more than $100\%$,~\cite{SAPT_Ions_Ka_Un_2015} the more sophisticated SAPT2+/aDZ method employed herein is expected to more accurately describe $\varepsilon_{\rm Disp}$ in the $700$ anion-$\pi$ complexes present in \NENCI.
As such, this SAPT level should be well-suited to provide a physically sound and semi-quantitative characterization of the binding motifs included in \NENCI.

In previously constructed databases of non-covalent interactions (\eg S66~\cite{rezac_s66_2011} and S101~\cite{wang_s101x7_2015}), each intermolecular complex was typically classified into one of three categories, based on whether $E_{\rm int} \approx \varepsilon_{\rm SAPT}$ was dominated by the $\varepsilon_{\rm Elst}$ component (Elst-bound), the $\varepsilon_{\rm Disp}$ component (Disp-bound), or a mixture (Mix) thereof (Mix-bound).
Since the $\varepsilon_{\rm Ind}$ component tended to be small in these complexes, the analogous and fourth Ind-bound category was deemed to be largely unnecessary.
With the addition of ${1,400}$ ion-$\pi$ complexes (in particular, the $700$ cation-$\pi$ systems), the scope of the SAPT decomposition analysis is substantially wider in \NENCI, and now encompasses the Ind-bound regime.~\cite{DNA_IonPi_2020}
As such, we propose a natural extension of the traditional three-category classification scheme made popular by Hobza \textit{et al.}~\cite{rezac_s66_2011} and Sherrill \textit{et al.}~\cite{Sherrill_Silver_Bronze_NCI_2014,sherrill_bfdb_2017} to include the Ind-bound category.
To do so, we construct a three-dimensional feature space defined by the $\varepsilon_{\rm Disp}/\varepsilon_{\rm Elst}$, $\varepsilon_{\rm Ind}/\varepsilon_{\rm Disp}$, and $\varepsilon_{\rm Elst}/\varepsilon_{\rm Ind}$ ratios as follows: \\[-0.5em]

\noindent \textbf{STEP 1.} To start, a single dimension of the feature space is chosen as the basis for constructing an initial sub-classification scheme.
Although this choice is arbitrary, we will start with the $\varepsilon_{\rm Disp}/\varepsilon_{\rm Elst}$ ratio, as this selection is tantamount to constructing the aforementioned three-category classification scheme (\ie Elst-bound, Disp-bound, or Mix-bound). 
For illustrative purposes, a ternary diagram ($T_{\rm ED}$) depicting this initial sub-classification scheme is plotted in the left panel of Fig.~\ref{fig:ternary}. \\[-0.5em]

\noindent \textbf{STEP 2.} Intermolecular complexes with $\left| \varepsilon_{\rm Disp}/\varepsilon_{\rm Elst} \right| > \eta$ are sub-classified as Disp-bound (shaded blue regions in $T_{\rm ED}$), while intermolecular complexes with $\left| \varepsilon_{\rm Elst}/\varepsilon_{\rm Disp} \right| > \eta$ are sub-classified as Elst-bound (shaded green regions in $T_{\rm ED}$).
If one stopped at this point, set $\eta = 2$, and classified all other cases as Mix-bound, this initial sub-classification scheme (based on the single $\varepsilon_{\rm Disp}/\varepsilon_{\rm Elst}$ feature) would be equivalent to the three-category classification scheme described above.
Since the value of $\eta$ is somewhat arbitrary, we have chosen to employ a slightly smaller value ($\eta = 3/2$) in the classification scheme introduced in this work; with this choice for $\eta$, less intermolecular complexes will be classified as Mix-bound (\textit{vide infra}). \\[-0.5em]

\noindent \textbf{STEP 3.} To go beyond this three-category classification scheme, STEP 2 is repeated for the two remaining dimensions of the feature space.
Selection of the $\varepsilon_{\rm Ind}/\varepsilon_{\rm Disp}$ feature generates the $T_{\rm ID}$ ternary diagram in Fig.~\ref{fig:ternary} and the analogous sub-classification of intermolecular complexes as: Disp-bound (if $\varepsilon_{\rm Disp}/\varepsilon_{\rm Ind} > \eta$; shaded blue regions in $T_{\rm ID}$) or Ind-bound (if $\varepsilon_{\rm Ind}/\varepsilon_{\rm Disp} > \eta$; shaded red regions).
Similarly, the $\varepsilon_{\rm Elst}/\varepsilon_{\rm Ind}$ feature yields the final required sub-classification scheme: Elst-bound (if $\left| \varepsilon_{\rm Elst}/\varepsilon_{\rm Ind} \right| > \eta$; shaded green regions in $T_{\rm EI}$) or Ind-bound (if $\left| \varepsilon_{\rm Ind}/\varepsilon_{\rm Elst} \right| > \eta$; shaded red regions).
Here, we note in passing that the absolute value (magnitude) must be used for all sub-classifications based on $\varepsilon_{\rm Elst}$, as the sign of the Elst component can be positive or negative. \\[-0.5em]

\noindent \textbf{STEP 4.} To arrive at our extended (\ie four-category) classification scheme, each intermolecular complex that has been sub-classified (in STEP 2 and STEP 3) with the same label twice retains that label; otherwise, the intermolecular complex is classified as Mix-bound.
This final classification scheme is graphically depicted in the colored $T_{\rm EID}$ ternary diagram in Fig.~\ref{fig:ternary}, which is assembled as an ``outer sum'' over the colored ternary diagrams corresponding to the sub-classification schemes, \ie $T_{\rm EID} = T_{\rm ED} \oplus T_{\rm ID} \oplus T_{\rm EI}$, in which the colors of $T_{\rm EID}$ are determined according to the rules described above. \\[-0.5em]

Based on this extended four-category classification scheme, the $141$ equilibrium intermolecular complexes in \NENCI are comprised of $54$ ($38.3\%$) Elst-bound, $23$ ($16.3\%$) Ind-bound, $31$ ($22.0\%$) Disp-bound, and $33$ ($23.4\%$) Mix-bound dimers.
When including all non-equilibrium intermolecular distances and angles, the entire \NENCI database contains ${3,499}$ ($45.1\%$) Elst-bound, $700$ ($9.0\%$) Ind-bound, ${1,372}$ ($17.7\%$) Disp-bound, and ${2,192}$ ($28.2\%$) Mix-bound intermolecular complexes.
Here, we note in passing that this observed decrease in the percentage of Ind-bound complexes is partially due to the inclusion of five (instead of nine) intermolecular angles for each ion-$\pi$ complex due to symmetry considerations (see Sec.~\ref{subsec:geometries}).
As such, the intermolecular complexes in \NENCI largely span the entire ternary diagram in Fig.~\ref{fig:ternary} and therefore contain a diverse array of binding motifs; as such, we hope that \NENCI will be used to critically examine (and potentially improve) the performance of theoretical models when faced with the challenge of simultaneously describing diverse NCI types on the same footing (\ie point (\textit{ii}) in the Introduction).

Here, we note that the apparent bias towards Elst-bound complexes in \NENCI is an unavoidable consequence of sampling short-range intermolecular separations; at such distances, there is often a substantial amount of orbital/density overlap between monomers, and charge penetration effects~\cite{Stone_Intermolecular_Forces,Sherrill_Nuc_Charge_Penetration_2015,Charge_Penetration_AMOEBA_2017,Sherrill_BZ_CP_2011,Novotny_LP-pi_analysis_and_bio_2016,DNA_IonPi_2020} (in $\varepsilon_{\rm Elst}$) tend to be the dominant contribution (over $\varepsilon_{\rm Ind}$ and $\varepsilon_{\rm Disp}$) to $\varepsilon_{\rm SAPT}$.
For instance, a significant majority ($73.9\%$) of the intermolecular complexes at $0.7\times$ are classified as Elst-bound while approximately half that ($38.3\%$) of the $141$ equilibrium dimers share this label.
This increase in the relative number of Elst-bound complexes at shorter intermolecular separations is clearly reflected in the ternary diagram in the right panel of Fig.~\ref{fig:ternary} as well as the percentage of Elst-bound complexes when broken down by the scaled intermolecular distance, \ie $73.9\%$ ($0.7\times$), $48.9\%$ ($0.8\times$), $40.5\%$ ($0.9\times$), $38.1\%$ ($0.95\times$), $37.3\%$ ($1.0\times$), $38.0\%$ ($1.05\times$), and $38.9\%$ ($1.1\times$).
In general, many complexes that are Disp-bound at larger intermolecular distances become Elst-bound or Mix-bound at reduced separations where short-range effects (\eg charge penetration) become more significant.
On the other hand, the Ind-bound complexes (which are primarily comprised of cation-$\pi$ interactions) tend to remain Ind-bound even at reduced intermolecular separations since charge penetration effects are substantially reduced when one of the monomers is a monovalent cation (\eg \ce{Li+} or \ce{Na+}).~\cite{DNA_IonPi_2020} 
For reference, the respective percentages of Ind-bound, Disp-bound, or Mix-bound complexes as a function of the scaled intermolecular distance are: $6.0\%$, $0.0\%$, $20.2\%$ for $0.7\times$, $7.6\%$, $0.0\%$, $43.6\%$ for $0.8\times$, $8.6\%$, $15.2\%$, $35.7\%$ for $0.9\times$, $10.0\%$, $22.8\%$, $29.0\%$ for $0.95\times$, $10.3\%$, $27.8\%$, $24.6\%$ for $1.0\times$, $10.3\%$, $28.7\%$, $23.1\%$ for $1.05\times$, and $10.5\%$, $29.2\%$, $21.5\%$ for $1.1\times$.

Before moving on to consider the error/uncertainty in the $E_{\rm int}$ values in \NENCI, we note in passing that the positive electrostatics (Elst(+)) region of the ternary diagram in Fig.~\ref{fig:ternary} is not sampled as well as the (Elst(-)) region.
However, \NENCI does contain a non-negligible ($413$) number of intermolecular complexes with $\varepsilon_{\rm Elst} > 0$. 
As mentioned above, such complexes are primarily found among the cation-$\pi$ complexes, where the degree of orbital overlap in the dimer (and hence the energetic stabilization due to charge penetration effects) is largely 
suppressed;~\cite{DNA_IonPi_2020} hence, intermolecular complexes with repulsive $\varepsilon_{\rm Elst}$ values are quite rare and may be adequately accounted for in \NENCI.

\subsection{Error Analysis and Critical Assessment of the Benchmark Intermolecular Interaction Energies \label{sec:benchmarkerror}}

\begin{figure*}[t!]
    \centering
    \includegraphics[width=0.95\textwidth]{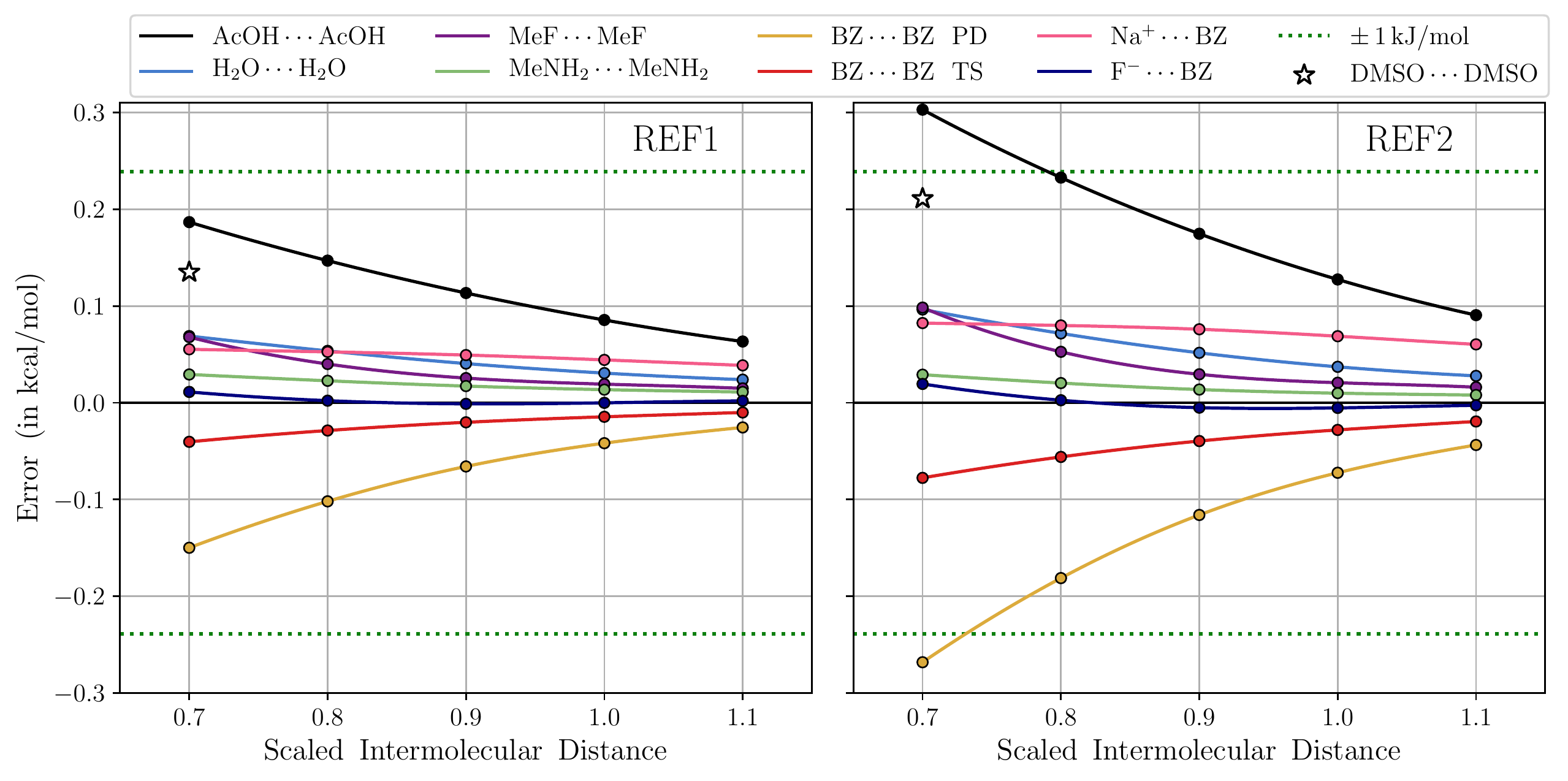}
    \caption{ 
    Errors (in kcal/mol) in the \NENCI $E_{\rm int}$ values (computed using the $E^{\rm CCSD(T)/CBS}$ extrapolation scheme in Eqs.~\eqref{eqn:CBSExtrap1}--\eqref{eqn:CBSExtrap4}) with respect to $E^{\rm REF1}$ (Eq.~\eqref{eqn:deltaQextrap}; \textit{left}) and $E^{\rm REF2}$ (Eq.~\eqref{eqn:CBSExtrapTQ}; \textit{right}) for a representative set of intermolecular complexes.
    Plotted as a function of the scaled intermolecular distance (with the $0.95\times$ and $1.05\times$ distances omitted for clarity), all intermolecular complexes (with the exception of DMSO$\cdots$DMSO, black stars) were kept at their equilibrium angle.
    Errors with respect to $E^{\rm REF1}$ and $E^{\rm REF2}$ were computed as $E^{\rm CCSD(T)/CBS}-E^{\rm REF1}$ and $E^{\rm CCSD(T)/CBS}-E^{\rm REF2}$, respectively.
    Based on this error profile (see main text), the errors in the \NENCI $E_{\rm int}$ values are $\pm \, 0.1$~kcal/mol on average, but can be as large as $\pm \, 0.2\mathrm{-}0.3$~kcal/mol (\ie $\pm \, 1$~kJ/mol, dashed green lines) for some complexes at reduced (\ie $0.7\times\mathrm{-}0.8\times$) intermolecular separations.
    }
    \label{fig:errorofbenchmark}
\end{figure*}

In addition to being extensive in size and scope, we would also argue that a well-balanced database of intermolecular interactions should contain a reliable estimate of the error/uncertainty present in the computed $E_{\rm int}$ values.
For \textit{ab initio} WFT methods, the two primary sources of error when computing $E_{\rm int}$ are: (\textit{i}) incompleteness in the one-particle basis set (\ie basis set incompleteness error (BSIE)) and (\textit{ii}) the approximate treatment of the electron correlation energy ($E_{\rm corr}$).
Since the mean-field HF contribution to $E_{\rm int}$ converges quickly with respect to the underlying basis set,~\cite{Halkier_basis_set_convergence_HF_1999} we expect that the BSIE at the $E_{\rm int}^{\rm HF/aQZ}$ level will be negligible when compared to the BSIE in the post-HF correlation energy contributions in Eq.~\eqref{eqn:CBSExtrap1}.
As depicted in Eq.~\eqref{eqn:CBSExtrap2}, the BSIE in the MP2 correlation energy is largely mitigated using the two-point extrapolation scheme~\cite{Helgaker_Basis_Set_Convergence_1998} for approximating the MP2/CBS limit provided in Eq.~\eqref{eqn:CBSExtrap3}. 
Although the $\delta E^{\rm CCSD(T)}$ correction converges with respect to the basis set significantly faster than $E_{\rm corr}^{\rm MP2}$ or $E_{\rm corr}^{\rm CCSD(T)}$ alone,~\cite{NCO_focal_point_1993,Sherrill_BZ_Dimer_Limit_2002,Sherrill_BZ_Dimer_2006,Hobza_delta_CC_Convergence_Stacks_2009} the BSIE in this term is generally the largest remaining source of error for extrapolation schemes such as that outlined in Eqs.~\eqref{eqn:CBSExtrap1}--\eqref{eqn:CBSExtrap4}.~\cite{Boese_CC_Basis_Set_2007,Sherrill_DeltaCC_extrapolation_2011}
To mitigate this error (and still remain computationally feasible when generating such a large number of intermolecular interaction energies), this contribution was computed using an augmented Dunning-style triple-$\zeta$ (haTZ) basis set~\cite{basis_Dunning_aXZ_H_Ne_1989,Dunning_Diffuse_aXZ_1992,basis_Dunning_aXZ_Al_Ar_1993,Dunning_aXZ_basis_Ga_Kr_1999} in \NENCI (\cf Eq.~\eqref{eqn:CBSExtrap4}).

As such, we will primarily focus on the remaining BSIE in the $\delta E^{\rm CCSD(T)/haTZ}$ contribution to $E_{\rm int}$ when critically assessing the accuracy of the intermolecular interaction energies in \NENCI.
To do so, we will compare our $E_{\rm int}$ values against two different references.
As a first reference value, we computed the $\delta E^{\rm CCSD(T)}$ correction in Eq.~\eqref{eqn:CBSExtrap4} using a larger (and substantially more expensive) augmented quadruple-$\zeta$ (aQZ) basis set, \ie
\begin{align}
    E^{\rm REF1} &= E^{\rm MP2/CBS} + \delta E^{\rm CCSD(T)/aQZ} \nonumber \\
                 &= E^{\rm HF/aQZ} + E_{\rm corr}^{\rm MP2/a(TQ)Z} + \delta E^{\rm CCSD(T)/aQZ} .
    \label{eqn:deltaQextrap}
\end{align}
in which $E^{\rm MP2/CBS}$ was computed using Eqs.~\eqref{eqn:CBSExtrap2}--\eqref{eqn:CBSExtrap3}.
As a second and alternative reference, we simply replaced the $\delta E^{\rm CCSD(T)/haTZ}$ correction with a direct two-point extrapolation~\cite{Helgaker_Basis_Set_Convergence_1998} of $E^{\rm CCSD(T)}$ using the aTZ and aQZ basis sets, \ie
\begin{align}
    E^{\rm REF2} &= E^{\rm CCSD(T)/a(TQ)Z} \nonumber \\
                 &= E^{\rm HF/aQZ} + E_{\rm corr}^{\rm CCSD(T)/a(TQ)Z} . 
    \label{eqn:CBSExtrapTQ}
\end{align}
By including CCSD(T) calculations in the much larger aQZ basis set, both of these reference values directly probe the BSIE in the CCSD(T) contribution, and are expected to be more reliable than the $E_{\rm int}$ values in the \NENCI database.

The error of the CCSD(T)/CBS scheme outlined in Eqs.~\eqref{eqn:CBSExtrap1}--\eqref{eqn:CBSExtrap4} with respect to both $E^{\rm REF1}$ and $E^{\rm REF2}$ is shown in Fig.~\ref{fig:errorofbenchmark} for a select subset of intermolecular complexes in \NENCI.
Plotted as a function of the scaled intermolecular distance (at the equilibrium angle, unless otherwise noted), this subset of intermolecular complexes was chosen to cover the wide array of binding motifs in \NENCI, and includes examples of Elst-, Ind-, Disp-, and Mix-bound systems, \ie single (\ce{H2O}$\cdots$\ce{H2O}, \ce{MeNH2}$\cdots$\ce{MeNH2}) and double (\ce{AcOH}$\cdots$\ce{AcOH}) hydrogen bonds, dipole-dipole (\ce{MeF}$\cdots$\ce{MeF}), $\pi$-$\pi$ stacking (BZ$\cdots$BZ PD), CH-$\pi$ (BZ$\cdots$BZ TS), as well as cation-$\pi$ (\ce{Na+}$\cdots$BZ) and anion-$\pi$ (\ce{F-}$\cdots$BZ) interactions.
As seen in Fig.~\ref{fig:errorofbenchmark}, the $E_{\rm int}$ values in \NENCI are generally within $\pm \, 0.1$~kcal/mol of both $E^{\rm REF1}$ and $E^{\rm REF2}$, and the errors with respect to these references tend to increase in magnitude at reduced intermolecular distances.
The worst-case scenarios among this subset include the acetic acid dimer (\ce{AcOH}$\cdots$\ce{AcOH}, double hydrogen-bonded) and the $C_{\rm 2h}$ parallel-displaced (PD) benzene dimer (BZ$\cdots$BZ PD, $\pi$-$\pi$ stacking), with errors in both steadily increasing in magnitude as the intermolecular separation is decreased; at $0.7 \times$, we report errors of $+0.19$~kcal/mol (\ce{AcOH}$\cdots$\ce{AcOH}) and $-0.15$~kcal/mol (BZ$\cdots$BZ PD) with respect to $E^{\rm REF1}$ ($+0.30$~kcal/mol and $-0.27$~kcal/mol when compared to $E^{\rm REF2}$, \textit{vide infra}). 
In these cases, the increased error is most likely due to the relatively larger amount of orbital overlap between these monomers at reduced intermolecular separations, where the interplay between short-range intermolecular interactions (\ie charge penetration, Pauli repulsion, many-body exchange-correlation effects, etc) becomes increasingly more challenging to describe in an accurate and reliable fashion.
This trend is also reflected in the error profiles corresponding to the two different BZ$\cdots$BZ dimers in Fig.~\ref{fig:errorofbenchmark}, where one can see that the error in the PD dimer (more orbital overlap) is noticeably larger in magnitude than the error in the $C_{\rm 2v}$ T-shaped (TS) dimer (less orbital overlap) at all intermolecular separations.
In the same breath, we also note that the error with respect to $E^{\rm REF1}$ (or $E^{\rm REF2}$) is non-trivial in general, and does not necessarily follow a direct/straightforward correlation with closest intermolecular contacts and/or the sign/magnitude of $E_{\rm int}$.
For example, the errors for \ce{AcOH}$\cdots$\ce{AcOH} ($E_{\rm int} = +12.1$~kcal/mol) and BZ$\cdots$BZ PD ($E_{\rm int} = +51.2$~kcal/mol) at $0.7\times$ are both larger than that found in the intermolecular complex with the largest (most repulsive) $E_{\rm int}$ value and closest \ce{O}$\cdots$\ce{H} distance in \NENCI---a non-equilibrium configuration of DMSO$\cdots$DMSO with $E_{\rm int} = +186.8$~kcal/mol (whose errors with respect to $E^{\rm REF1}$ and $E^{\rm REF2}$ are depicted by stars in Fig.~\ref{fig:errorofbenchmark}).

From this analysis, we believe that the errors in the \NENCI $E_{\rm int}$ values are mostly within $\pm \, 0.1$~kcal/mol, but can be as large as $0.2\mathrm{-}0.3$~kcal/mol (\ie $\approx 1$~kJ/mol) for certain systems at reduced intermolecular separations.
Here, we note in passing that the $\delta E^{\rm CCSD(T)/haTZ}$ correction used in \NENCI provides a significant improvement over $\delta E^{\rm CCSD(T)/aDZ}$, and yields nearly identical $E_{\rm int}$ values when compared to the more expensive $\delta E^{\rm CCSD(T)/aTZ}$ approach; this is shown in Fig.~S1 and again emphasizes the need for triple-$\zeta$ basis sets when employing the $\delta E^{\rm CCSD(T)}$ correction scheme.~\cite{Sherrill_DeltaCC_extrapolation_2011}
When considering the largest errors in Fig.~\ref{fig:errorofbenchmark}, \ie \ce{AcOH}$\cdots$\ce{AcOH} and BZ$\cdots$BZ PD, one can see that the errors with respect to $E^{\rm REF1}$ and $E^{\rm REF2}$ differ by $\approx 0.1$~kcal/mol; as such, the estimated \textit{average} error in \NENCI ($\pm \, 0.1$~kcal/mol) is comparable to the difference between using $E^{\rm REF1}$ or $E^{\rm REF2}$ as the reference for $E_{\rm int}$.
Generally speaking, it is not clear which of these two quantities supplies the more accurate reference for $E_{\rm int}$; however, it has been pointed out by Sherrill and co-workers~\cite{Sherrill_DeltaCC_extrapolation_2011} that the $\delta E^{\rm CCSD(T)}$ correction does not converge monotonically towards the CBS limit, which implies that $E^{\rm REF1}$ might in fact be a slightly better reference value than $E^{\rm REF2}$.

As mentioned above, the other primary source of error when computing $E_{\rm int}$ using approximate \textit{ab initio} WFT methods is the necessarily incomplete treatment of the electron correlation energy; while post-CCSD(T) corrections tend to be small for \textit{equilibrium} intermolecular interaction energies (\ie $< 0.1$~kcal/mol),~\cite{Hobza_ccsdt_gold_standard_2013} whether or not such corrections become more substantial at reduced intermolecular separations still remains unanswered.
With increasingly unfavorable scaling with both system and basis set size, such post-CCSD(T) calculations (\ie CCSDT, CCSDT(Q), CCSDTQ, etc) are computationally prohibitive and could have only been performed on: (\textit{i}) the smaller/smallest systems in \NENCI, but with sufficiently large basis sets (of at least triple-$\zeta$ or quadruple-$\zeta$ quality) or (\textit{ii}) the larger/largest systems in \NENCI, but with reduced and insufficiently large basis sets (\ie double-$\zeta$ at best). 
Since neither of these approaches would have provided an accurate and reliable estimate of the post-CCSD(T) contributions to $E_{\rm int}$ for the wide range of intermolecular complexes in \NENCI,~\cite{Hobza_composite_CCSDTpQ_schemes_2014} we chose to focus our efforts above on critically assessing the CCSD(T)/CBS scheme outlined in Eqs.~\eqref{eqn:CBSExtrap1}--\eqref{eqn:CBSExtrap4} based on a quantitative estimate of the remaining BSIE at the CCSD(T) level.
Since an accurate and reliable prediction of $E_{\rm int}$ for intermolecular complexes in the repulsive wall (\ie inside the vdW envelope) poses a substantive challenge to state-of-the-art DFT and WFT methods (see \pII~\cite{NENCI_part_II} in this series), further benchmarking of the standard CCSD(T)/CBS approach (possibly \via stochastic CC~\cite{Piecuch_CC_Monte_Carlo_2017,Crawford_Thom_Diagramatic_CC_Monte_Carlo_2019,Crawford_Thom_Stochastic_CC_2020} or FCI~\cite{alavi_stochastic_FCI_MC_2015,Umrigar_Stochastic_PT_to_CI_2017,Umrigar_Stochastic_CI_2018} methods) in this regime is an open challenge for the community and will be of critical importance for the development of next-generation DFT functionals and ML-based intra-/inter-molecular interaction potentials.

\section{Conclusions and Future Directions}

In this work, we present \NENCI: a large and comprehensive database of approximately ${8,000}$ benchmark non-equilibrium non-covalent interaction energies for a diverse selection of intermolecular complexes of biological and chemical relevance with a particular emphasis on close intermolecular contacts.
Designed to address the growing need for extensive high-quality quantum mechanical data in the chemical sciences, \NENCI starts with the $101$ molecular dimers in the widely used S66, S66x8, S66a8, and S101x7 databases,~\cite{rezac_s66_2011,rezac_s66a8_2011,wang_s101x7_2015} and extends the scope of these popular works in two directions.
For one, \NENCI includes $40$ cation- and anion-$\pi$ complexes, a fundamentally important class of NCIs that are found throughout nature and among the strongest NCIs known.
Secondly, \NENCI systematically samples both equilibrium and non-equilibrium configurations on all $141$ intermolecular PES by \textit{simultaneously} varying the intermolecular distance (from $0.7\times\mathrm{-}1.1\times$ the equilibrium separation) and intermolecular angle (including either five or nine angles for each distance, depending on symmetry considerations). 
As such, a wide range of intermolecular atom-pair distances are present in \NENCI, including a large number of close intermolecular contacts with atom pairs located inside their respective vdW envelope; these intermolecular complexes probe a number of different short-ranged NCIs (\eg charge transfer and penetration, Pauli repulsion, many-body exchange-correlation effects, etc), which are observed in many important chemical and biological systems, and pose an enormous challenge for molecular modeling.
Computed at the CCSD(T)/CBS level of theory, the ${7,763}$ benchmark $E_{\rm int}$ values in \NENCI range from $-38.5$~kcal/mol (most attractive) to $+186.8$~kcal/mol (most repulsive), with a total span of $225.3$~kcal/mol and a mean (median) $E_{\rm int}$ value of $-1.06$~kcal/mol ($-2.39$~kcal/mol).
A detailed SAPT-based analysis was used to confirm the diverse and comprehensive nature of the intermolecular binding motifs present in \NENCI, which includes a significant number of primarily induction-bound dimers and now spans all regions of the SAPT ternary diagram; this warranted a new four-category classification scheme that includes complexes primarily bound by electrostatics (${3,499}$), induction ($700$), dispersion (${1,372}$), or mixtures thereof (${2,192}$).
Finally, a critical error analysis was performed on a representative set of intermolecular complexes, from which we estimate that the $E_{\rm int}$ values in \NENCI have a mean error of $\pm \, 0.1$~kcal/mol and a maximum error of $\pm \, 0.2\mathrm{-}0.3$~kcal/mol for the most challenging cases.

For all of these reasons, we believe that the \NENCI database is timely and well-suited for testing, training, and developing next-generation force fields, DFT and WFT methods, as well as ML based potentials.
An order-of-magnitude larger than any database of non-covalent interactions currently available, \NENCI can be used for a variety of different purposes.
For one, \NENCI could be employed as a single database and used in its entirety.
Alternatively, \NENCI can be split into multiple different training and testing data sets---each containing a diverse sample of intermolecular binding motifs---and used for cross-validation studies and statistical error assessment. 
When used for such purposes, we note in passing that strong  correlations will likely exist between different points on a given intermolecular PES; as such, we caution against separating such points between training and testing data sets to avoid issues associated with overfitting.

We end this manuscript with a brief discussion of several future research directions that could build off this work and potentially have an immediate impact in the field.
For one, \pII~\cite{NENCI_part_II} in this series (in preparation) will critically assess the accuracy and reliability of a large number of popular DFT and WFT methods when describing the diverse array of non-equilibrium non-covalent interactions in \NENCI, thereby identifying the strengths and weaknesses of established first-principles methods.
A simple and straightforward extension of \NENCI would target dimers with increased intermolecular distances (\eg beyond $1.1\times$ the equilibrium separation), as benchmark $E_{\rm int}$ values for such complexes could play an important role in testing and training ML methods for predicting molecular multipoles~\cite{veit2020predicting} and polarizabilities~\cite{yang2019quantum,alpha_ML_2019}, as well as addressing important unresolved questions regarding the treatment of long-range electrostatics in ML-based potentials.~\cite{yue2021short}
Other important research thrusts would focus on expanding \NENCI to further address the three challenges introduced above: (\textit{i}) the need to describe NCIs in large molecular and condensed-phase systems can be addressed with extensions that focus on large/complex systems and potentially include explicit solvent molecules; (\textit{ii}) the need to describe the diverse types of NCIs on the same footing can be addressed by including NCI binding motifs that are either not found or underrepresented in \NENCI (\eg triple hydrogen bonds, quadrupole-quadrupole interactions, ionic bonds, etc); (\textit{iii}) the need to describe NCIs in equilibrium and non-equilibrium systems on the same footing can be addressed by including complexes at more extreme (reduced and increased) intermolecular separations and angles as well as complexes between monomers in non-equilibrium configurations.

\begin{acknowledgments}
All authors thank David Sherrill for helpful scientific discussions and Destiny Malloy for creating an early prototype of Fig.~1.
All authors acknowledge partial financial support from Cornell University through start-up funding.
This material is based upon work supported by the National Science Foundation under Grant No. CHE-1945676.
RAD also gratefully acknowledges financial support from an Alfred P. Sloan Research Fellowship.
This research used resources of the National Energy Research Scientific Computing Center, which is supported by the Office of Science of the U.S. Department of Energy under Contract No. DE-AC02-05CH11231.
\end{acknowledgments}

\section*{Data Availability Statement}

The data that supports the findings of this study are available within the article and its supplementary material.

\section*{References}


%
%
%
%
%

\end{document}